\documentclass[12 pt]{article}

\usepackage[T1]{fontenc}

\usepackage{amsfonts,mathtools, amsmath,amsthm,amssymb,mhequ,dsfont, accents}

\usepackage{graphicx,booktabs,xcolor,rotating,float}

\usepackage[colorlinks,citecolor=gr,linkcolor=gr,urlcolor=gr]{hyperref}

\definecolor{prp}{HTML}{b16286}
\definecolor{bl}{HTML}{4E09B2}
\definecolor{gr}{HTML}{61635F}

\usepackage{titling}
\usepackage[top = 1.5cm , bottom= 2cm , left= 2cm , right= 2cm]{geometry}

\usepackage{setspace}
\usepackage{authblk}

\usepackage{titlesec}
\titleformat{\section}
  {\centering\bfseries}
  {\thesection}
  {.5em}
  {\MakeUppercase}

\titleformat{\subsection}
  {\normalfont\bfseries\itshape}{\thesubsection}{1em}{}

\usepackage[shortlabels]{enumitem}
\usepackage[font=small,labelfont=bf]{caption}

\usepackage{mdframed}
\mdfdefinestyle{st}{%
    linecolor=bl,
   linewidth=2pt,
        }

\usepackage[ruled,norelsize,vlined]{algorithm2e}
\usepackage[norelsize,vlined]{algorithm2e}
\usepackage[noabbrev,capitalise]{cleveref}
 \AtBeginDocument{%
\Crefname{algocf}{algorithm}{algorithms}
    \Crefname{section}{section}{sections}%
    \Crefname{figure}{Figure}{Figures}%
}

\usepackage{tikz}
\tikzstyle{none}=[inner sep=0mm]
\tikzstyle{nero}=[fill=black, draw=black, shape=circle]
\tikzstyle{rosso}=[-, fill=red, draw=red, thick]
\usetikzlibrary{backgrounds}
\usetikzlibrary{arrows}
\usetikzlibrary{shapes,shapes.geometric,shapes.misc}
\usetikzlibrary{arrows.meta, positioning}

\pgfdeclarelayer{edgelayer}
\pgfdeclarelayer{nodelayer}
\pgfsetlayers{background,edgelayer,nodelayer,main}

\newcommand{\argmin}{\operatornamewithlimits{argmin}}

\renewcommand\intercal{\mathsf{\scriptscriptstyle T}}

\newcommand{\T}{ {^\intercal} }
\newcommand{\bx}{ {\bf x} }
\newcommand{\bX}{ {\bf X} }

\newcommand{\bD}{ {\bf D} }
\newcommand{\by}{ {\bf y} }

\newcommand{\bxit}{{\bf x}_i\T }
\newcommand{\bxii}{{\bf x}_i}
\newcommand{\bQ}{ {\bf Q} }
\newcommand{\bS}{ {\bf S} } 

\newcommand{\br}{ {\bf r} }
\newcommand{\bV}{ {\bf V} }
\newcommand{\bW}{ {\bf W} }
\newcommand{\bU}{ {\bf U} }
\newcommand{\sumin}{\sum_{i=1}^n}

\newcommand{\mi}{_{\setminus i}}
\newcommand{\Qmi}{\bQ\mi}
\newcommand{\Qmiinv}{\bQ\mi^{-1}}
\newcommand{\Smi}{\bS\mi}
\newcommand{\rmi}{\br\mi}
\newcommand{\ep}{_{\textsc{ep}}}

\newcommand{\mf}{_{\textsc{mf}}}
\newcommand{\pmf}{_{\textsc{pmf}}}
\newcommand{\new}{_{\mbox{\scriptsize \textsc{new}}}}
\newcommand{\Qep}{ {\bf Q}\ep}
\newcommand{\rep}{ {\bf r}\ep}
\newcommand{\Sep}{ {\bf S}\ep}
\newcommand{\papp}{ \undertilde{p}}

\newcommand{\bDelta}{ {\boldsymbol \Delta} }
\newcommand{\bbeta}{ {\boldsymbol \beta} }
\newcommand{\bmu}{ {\boldsymbol \mu} }
\newcommand{\bSigma}{ {\boldsymbol \Sigma} }

\newcommand{\bOmega}{ {\boldsymbol \Omega} }

\newcommand{\balpha}{ {\boldsymbol \alpha} }
\newcommand{\bxi}{ {\boldsymbol \xi} }
\newcommand{\bUpsilon}{ {\boldsymbol \Upsilon} }
\newcommand{\bz}{ {\bf z} }

\newcommand{\repo}{\url{github.com/emanuelealiverti/epcp}}

\usepackage{natbib}

\title{Approximate Bayesian inference for cumulative probit regression models}
\author{Emanuele Aliverti}
\affil{Department of Statistical Sciences,  University of Padova}
\date{}

\begin{document}
\maketitle
\vspace{-2cm}
\begin{abstract}
	Ordinal categorical data are routinely encountered in many practical applications. 
	When the primary goal is to construct a regression model for ordinal outcomes, cumulative link models represent one of the most popular choices to link the cumulative probabilities of the response with a set of covariates through a parsimonious linear predictor, shared across response categories.
	As the number of observations grows, standard sampling algorithms for Bayesian inference scale poorly, making posterior computation increasingly challenging for large datasets. 
	In this article, we propose three scalable algorithms for approximating the posterior distribution of the regression coefficients in cumulative probit models relying on Variational Bayes and Expectation Propagation.
	We compare the proposed approaches with inference based on Markov Chain Monte Carlo, demonstrating superior computational performance and remarkable accuracy. 
	Finally, we illustrate the utility of the proposed algorithms on a challenging case study to investigate the structure of a criminal network.
\end{abstract}
\section{Introduction}
\label{sec:intro}
In many scientific disciplines, ordinal categorical variables are frequently collected to measure outcomes of interest. 
For example, ordinal variables are routinely employed in the social sciences to capture attitudes and opinions using Likert scales, allowing respondents to express their level of agreement with various statements through ordered categories such as ``strongly disagree,'' ``disagree,'' ``undecided,'' ``agree,'' and ``strongly agree'' \citep{agresti:2010}. 
Other common examples include measuring how frequently specific activities are performed, the severity of symptoms in clinical or medical assessments, or the level of satisfaction with a service \citep[e.g.,][]{gambarota:2024,varin:2010,gill:2009}; ordinal data naturally arise in diverse domains, making them pervasive across many areas of empirical research.
When ordinal variables are used as response outcomes in regression modeling, appropriate statistical models that properly account for their nature are required; for example, treating ordinal responses as nominal ignores the clear ordering of their levels and typically results in overparameterized models, while considering them as continuous implies equal spacing between categories, an assumption that is rarely justifiable in practice \citep{agresti:2010}.

This article focuses on the cumulative probit model \citep{mccullagh:1980}, also known as {\em ordered probit} model \citep[Section 5.2]{agresti:2010}, one of the most popular approaches for ordinal regression.
The cumulative probit model belongs to the broader class of {\em cumulative link models}, characterized by a common linear predictor shared across all response categories; according to this assumption, the effect of the covariates on the response is the same for each response level, providing a parsimonious---yet flexible---framework for modeling ordinal outcomes \citep{agresti:2010}. 
Specifically, in the cumulative probit model we denote for each statistical unit $i=1,\dots,n$ the ordinal response value as $y_i \in\{1, \dots, K\}$, and with $\bx_i = (x_{i1}, \dots, x_{ip})\T \in \mathbb{R}^p$ a vector of $p$ covariates; the cumulative probabilities of the ordinal response variable are linked with the covariates as
	\begin{equs}
		\label{eq:cum}
		\mbox{pr}(y_i \leq k) = \Phi(\alpha_k - \bxit\bbeta), \quad k = 1, \dots, K-1,
\end{equs}
where $\alpha_0 < \alpha_1 < \dots < \alpha_{K}$ are ordered cutpoints (or thresholds), $\bbeta = (\beta_1, \dots, \beta_p)\T$ denotes a vector of regression coefficients and $\Phi(\cdot)$ the cumulative distribution function (\textsc{cdf}) of a standard Gaussian.
\begin{figure}
\centering
\fbox{\includegraphics[width=.5\textwidth]{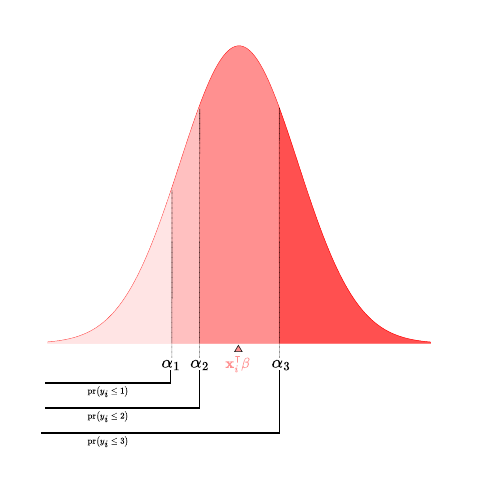}}
\caption{Construction of cumulative probit probabilities. The red curve represents the density of the latent $z_i$, distributed as a Gaussian with mean $\bxit\bbeta$ and unit variance. Cumulative probabilities are obtained by discretizing such a distribution as in~\eqref{eq:cum}; colored regions correspond to the probabilities of distinct categories, defined in~\eqref{eq:probs}.}
\label{fig:ordinal}
\end{figure}

Cumulative link models can be interpreted as arising from a discretized latent regression formulation, where the observed ordinal response $y_i$ indicates the interval in which an unobserved continuous latent variable $z_i$ falls; in the cumulative probit, such a latent response follows a Gaussian linear regression model with mean $\bxit\bbeta$ and unit variance, and the observed value $y_i$ is equal to $k$ if $\alpha_{k-1} \leq z_i < \alpha_k$, for each $k = 1, \dots, K$ \citep[][Section 4.1]{albert:1993}; refer also to Figure~\ref{fig:ordinal} for a graphical illustration of this discretization process.
Such a representation highlights that the distances between adjacent ordered factor levels are not constant, but are instead regulated by the cutpoints; moreover, it suggests to interpret the regression coefficients as effects on the latent scale---that is, as {increases} in the expected value of the latent response corresponding to a one-unit {increase} in a given covariate, adjusting for other predictors \citep[e.g.][Section 5.1.3]{agresti:2010}. 
This representation justifies the common practice to parametrize the regression coefficients in~\eqref{eq:cum} with a negative sign, in order to make their interpretation in the latent scale coherent with the usual directional interpretation \citep[e.g.,][]{scarpa:2012}.

Under a Bayesian approach, posterior inference for the cumulative probit model generally relies on sampling via Markov Chain Monte Carlo (\textsc{mcmc}), leveraging its latent variable representations \citep{albert:1993,albert:2001} or relying on Hamiltonian Monte Carlo \citep[e.g.,][]{brms}; refer also to \citet{jeliazkov:2008} for an extensive review.
However, as the number of observations $n$ increases, posterior sampling rapidly becomes computationally demanding for practical applications, thereby motivating the development of more scalable routines that can still deliver accurate results. 
While the widespread use of ordinal regression models has recently fostered methodological advancements in scalable maximum likelihood estimation \citep[e.g.,][]{ordinal,bellio:2025,gronberg:2025}, less attention has been devoted to the accurate approximation of the posterior distribution of the ordinal probit under a Bayesian framework.

In this article, we address this gap by proposing a general framework for approximate posterior inference under the cumulative probit model, introducing three complementary algorithmic routines based on variational inference \citep[e.g.,][Chapter~10]{bishop:2006}. 
The first two algorithms rely on Mean-Field Variational Bayes \citep[e.g.,][]{ormerod:2010}, a broad class of techniques that approximate the posterior distribution with a tractable density subject to factorization constraints.
Both methods build upon the latent variable representation of the cumulative probit model discussed above, and use optimization to identify the best approximate density in the pre-specified tractable class by minimizing the forward Kullback–Leibler divergence.
The first algorithm approximates the joint posterior of the regression coefficients and latent variables using a fully factorized density, assuming independence between these two parameter blocks; the resulting algorithmic routine leads to satisfactory performance and serves as a fundamental building block for more structured derivations.
The second routine extends the partially factorized Mean-Field approach of \citet{fasano:2022} to the ordinal setting, factorizing the joint posterior as the product of the density for the latent variables and the conditional density of the regression coefficients given the latent variables; this broader variational family yields improved approximation accuracy while maintaining comparable computational cost.

The third proposal consists of an innovative routine based on Expectation Propagation \citep{opper:2000,minkaEP}, a framework for approximate inference that relies on the reverse Kullback–Leibler divergence. 
In this approach, the posterior distribution of the regression coefficients is replaced by a product of tractable factors that preserve the multiplicative structure of the exact posterior, and the resulting approximation is refined by iteratively improving each approximate factor.
Notably, the proposed routine for the cumulative probit is derived through probabilistic arguments based on the Selection-Normal distribution \citep{arellano:2006}, leading to a neat implementation that involves univariate Truncated-Normal moments and rank-one matrix operations.
This result is particularly remarkable, as the derivation of Expectation Propagation algorithms often entails substantial algebraic and computational overhead \citep[e.g.,][]{kim:2016,kim2018expectation,hall2020fast}; instead, the proposed routine is analytically uncluttered, computationally efficient, and crucially leads to an approximation with very accurate empirical performance in most settings.

Despite its apparent simplicity, the cumulative model with a linear predictor as in~\eqref{eq:cum} encompasses a large class of regression models for ordinal responses that are routinely used in practice. 
Indeed, the combination of such a specification with Gaussian priors on the regression coefficients allows one to recast, within a unified framework, several random-effects models with generalized designs \citep{zhao:2006}; notable examples include models with random intercepts and slopes \citep[e.g.,][]{pinheiro:2000}, semiparametric regression via penalized splines \citep[e.g.,][]{ruppert:2003}, longitudinal models \citep[e.g.,][]{diggle:2002}, and additive social-regression models for network data (\citealp{hoff:2021}; see also Section~\ref{sec:criminal}). 
As a consequence, performing scalable inference via the proposed routines yields substantial benefits for several model specifications, providing a flexible and computationally efficient framework applicable to a wide range of statistical problems. 
In addition to providing inference for the regression coefficients, the proposed approximations also allow for tractable computation of other functionals such as predictive probabilities and the marginal likelihood, key-quantities for classification of future observations, estimation of hyper-parameters and model selection.
To the best of our knowledge, such a comprehensive effort to develop, compare, and unify multiple approximate routines for the cumulative probit model within a single coherent framework is currently lacking.
The code implementing the proposed methods in \texttt{c++} and providing an \texttt{R} interface is available at the repository \repo.

\section{Computational methods for cumulative probit regression}
\label{sec:comp}
\subsection{Model specification}
\label{sec:model}
Under the cumulative probit model, the multinomial likelihood function for a single observation $y_i\in\{1,\ldots,K\}$ can be derived by transforming the cumulative probabilities \eqref{eq:cum} into cell probabilities, letting for each statistical unit $i=1,\ldots,n$ and response level $k = 1, \dots, K$
\begin{equs}[e:block]
		\label{eq:probs}
		\pi_k(\bx_i) := \mbox{pr}(y_i = k) &= \mbox{pr}(y_i \leq k) - \mbox{pr}(y_i < k)\\ 
&= \mbox{pr}(y_i \leq k) - \mbox{pr}(y_i \leq k-1)\\ 
		 &= \Phi(\alpha_{k} -\bxit\bbeta) - \Phi(\alpha_{k-1} -\bxit\bbeta),
\end{equs}
where $\alpha_0 = -\infty$ and $\alpha_K = \infty$ \citep[e.g.,][Section 3.2]{agresti:2010}.
Under a Bayesian approach, it is common to assign to the regression coefficients a multivariate Gaussian prior with mean vector $\bmu_0$ and covariance $\bSigma_0$, and the associated prior density is denoted as $p_0(\bbeta) = \phi_p(\bbeta - \bmu_0, \bSigma_0)$. The cumulative probit model can be fully specified letting for each $i=1,\dots,n$
\begin{equs}[e:block]
	\label{eq:model}
	\bbeta &\sim \mbox{N}_p(\bmu_0, \bSigma_0)\\
\multicol{2}{\pi_k(\bx_i) := \Phi(\alpha_{k} -\bxit\bbeta) - \Phi(\alpha_{k-1} -\bxit\bbeta),  \quad k= 1, \dots, K}\\
	(y_i \mid \pi_1(\bx_i), \dots, \pi_K(\bx_i)) &\sim \mbox{Multinom}(\pi_1(\bx_i), \dots, \pi_K(\bx_i)),
\end{equs}
and denoting as $\ell_i(\bbeta; \balpha)$ the likelihood function for a single data point $(y_i, \bxit)$, it holds that
\begin{equs}[e:block]
		\label{eq:likelihood}
		\ell_i(\bbeta; \balpha) &= \prod_{k=1}^K \left[\Phi(\alpha_{k} -\bxit\bbeta) - \Phi(\alpha_{k-1} -\bxit\bbeta)\right]^{\mathds{I}[y_i = k]} \\
					&= \Phi(\alpha_{y_i} -\bxit\bbeta) - \Phi(\alpha_{y_i-1} -\bxit\bbeta),
\end{equs}
where $\mathds{I}[\mathcal{A}]$ denotes the indicator function of the event $\mathcal{A}$. The cutoffs $\balpha = (\alpha_0, \dots, \alpha_{K})\T$ act as intercept terms and are often considered as quantities of ``little interest'' \citep{mccullagh:1980}; for this reason, we treat them as nuisance parameters and initially focus on estimating the regression coefficients $\bbeta$ under fixed, generic values of $\balpha$. 
In Section~\eqref{sec:alphaopt}, we extend the proposed routines by introducing an Empirical Bayes strategy that optimizes the thresholds via (approximate) maximum marginal likelihood, thereby providing point estimates for the cutoffs in addition to the full posterior distribution of the regression coefficients.
In light of this, it is convenient to suppress the explicit dependence on $\balpha$ from the notation of the likelihood and related quantities; Equation~\eqref{eq:likelihood} is rewritten as $\ell_i(\bbeta)$ and the posterior distribution for the regression coefficients is written as
\begin{equs}[e:block]
	\label{eq:posterior}
	p(\bbeta \mid \by) = \frac{1}{p(\by)} \, {p_0(\bbeta)\prod_{i=1}^n \ell_i(\bbeta)},
\end{equs}
where $p(\by)$ indicates the marginal likelihood (or model evidence).

Although the cutoffs could also be regarded as additional parameters of interest and estimated within the proposed computational framework, the proposed strategy substantially simplifies the algorithmic derivations.
For example, under the data augmentation of \citet{albert:1993}---used within the derivations of Sections~\ref{sec:mfvb}~and~\ref{sec:pmf}---the full conditional distribution of each cutoff depends on the ordered statistics of the local augmented data, thereby increasing algebraic complexity and computational cost of the proposed derivations.
The implications of this choice in terms of underestimation of posterior uncertainty are assessed empirically in Section~\cref{sec:sim}, providing evidence that supports this modeling decision.

\subsection{Mean-field variational inference}
\label{sec:mfvb}
The focus of Mean-Field Variational Bayes (\textsc{mfvb}) is to approximate the posterior distribution with a density in a family that satisfies certain product-form restrictions \citep[e.g.,][Section 2.2]{ormerod:2010}.
Similarly to several discrete response models \citep[e.g.,][]{anceschi:2023}, the latent representation of the cumulative probit induces a conditionally conjugate construction with global variables $\bbeta$ and local augmented data $\bz = (z_1, \dots, z_n)\T$, recalling that model~\eqref{eq:model} can be obtained by integrating out $\bz$ from 
\begin{equs}
	\label{eq:mod_aug}
	\bbeta \sim \mbox{N}_p(\bmu_0, \bSigma_0), \quad	(z_i \mid \bbeta) \sim \mbox{N}(\bxit\bbeta,1), \quad (y_i\mid z_i) :=  \mathds{I}[\alpha_{k-1} \leq z_i < \alpha_k]\cdot k.
\end{equs}

The joint posterior $p(\bbeta,\bz \,|\,\by)$ is approximated with the closest member, in Kullback–Leibler (\textsc{kl}) sense, among all densities in the class ${\mathcal{Q}\mf= \left\{q\mf(\bbeta,\bz) :q\mf(\bbeta,\bz) = q\mf(\bbeta)q\mf(\bz) \right\}}$,
and the resulting approximation corresponds to
\begin{equs}[e:block]
\label{eq:klvb}
q^\star\mf(\bbeta,\bz) &=  \argmin_{q\mf(\bbeta,\bz) \in \mathcal{Q}\mf} {\textsc{kl}}\left[q\mf(\bbeta, \bz) \,||\, p(\bbeta,\bz\mid\by) \right]  .
\end{equs}
In practice, it is more convenient to maximize the alternative objective function
\begin{equs}[e:block]
\label{eq:elbo_def}
\log \papp(\by, q\mf) &:= \mathbb{E}_{q\mf(\bbeta,\bz)}[\log p(\by,\bz,\bbeta)] - \mathbb{E}_{q\mf(\bbeta,\bz)}[\log q\mf(\bbeta,\bz)] \\
			   &= -  {\textsc{kl}}\left[q\mf(\bbeta, \bz) \,||\, p(\bbeta,\bz\mid\by) \right]  + \log p(\by),
\end{equs}
which is often referred to as the {\em evidence lower bound} (\textsc{elbo}) since $\log \papp(\by, q\mf) \leq \log p(\by)$ for any $q\mf$ \citep[e.g.,][]{ormerod:2010}.
Maximizing the \textsc{elbo} is equivalent to minimizing the \textsc{kl} divergence, and such a maximization can be solved via an iterative scheme recalling that the optimal densities $q\mf^\star(\bbeta)$ and $q\mf^\star(\bz)$ satisfy
\begin{equs}[e:block]
\label{eq:cavi}
q\mf^\star(\bbeta) \propto \exp\left\{\mathbb{E}_{q\mf(\bz)} \log p(\bbeta \mid \bz,\by)\right\}, \quad q\mf^\star(\bz) \propto \exp\left\{\mathbb{E}_{q\mf(\bbeta)} \log p(\bz \mid \bbeta, \by)\right\},
\end{equs}
and updating iteratively each density using the current estimates for all of the other factors until changes in $\log \papp(\by,q\mf)$ are negligible \citep[e.g.,][Section 2.2]{bishop:2006, ormerod:2010}. 

Under the latent variable representation \eqref{eq:mod_aug}, the required full-conditional distributions are
\begin{equs}[e:block]
\label{eq:full_cond}
	(\bbeta \mid \bz, \by) &\sim \mbox{N}_p(\bV( \bSigma_0^{-1}\bmu_0+\bX\T\bz ), \bV), \quad &\bV=(\bSigma_0^{-1} + \bX\T\bX)^{-1}\\
(z_i \mid \bbeta, \by) &\sim \mbox{TN}([\alpha_{y_i-1}, \alpha_{y_i}], \bxit\bbeta,1), \quad &i=1,\dots, n,
\end{equs}
where $\bX$ indicates the design matrix with rows $\bxit$ and $\mbox{TN}([\alpha_{y_i-1}, \alpha_{y_i}], \bxit\bbeta,1)$ the density of a Truncated-Normal with location $\bxit\bbeta$ and unit scale restricted over the interval $[\alpha_{y_i-1}, \alpha_{y_i}]$.
The conditional independence structure of~\eqref{eq:mod_aug} implies that $p(\bz\mid\bbeta,\by) = \prod_{i=1}^n p(z_i\mid\bbeta,\by)$, and such a structure is induced in the family $\mathcal{Q}\mf$ that further factorizes as $q\mf(\bbeta)q\mf(\bz)=q\mf(\bbeta)\prod_{i=1}^n q\mf(z_i)$; combining this factorization with Equations~\eqref{eq:cavi} and \eqref{eq:full_cond}, the optimal variational densities $q\mf(\bbeta)$ and $q\mf(z_i)$ are in the same Gaussian and Truncated-Normal families of their full-conditionals, with parameters that are iteratively updated via variational expectations until convergence.

Algorithm~\ref{algoMFVB} illustrates the iterative updates that characterize the \textsc{mfvb} procedure, relying on the function $\zeta_1(a,b) = [\phi(b) - \phi(a)][\Phi(b) - \Phi(a)]^{-1}$ for computing expected values of univariate Truncated-Normals. 
A numerically stable implementation of such function is available at the repository \repo, generalizing the approach suggested in \citet{sn-pack} to the bilateral case and using the asymptotic expansion (26.2.13) of \citet{Abramowitz1964} for very large values of $|a|$ or $|b|$.
The \textsc{elbo} can be explicitly computed at the end of each iteration as
\begin{equs}[e:block]
\label{eq:elbo_full}
\log \papp(\by, q\mf) &= \sumin \log[{\Phi(\alpha_{y_i} -\bxit\bar{\bbeta}) - \Phi(\alpha_{{y_i} - 1} -\bxit\bar{\bbeta})}] -\frac{1}{2} (\bar{\bbeta}-\bmu_0)\T\bSigma_0^{-1}(\bar{\bbeta}-\bmu_0) + const\quad\quad
\end{equs}
where ``${const}$'' denotes an additive constant not depending on the variational parameters; full derivations are reported in Section~\ref{sec:suppMF} of the Supplementary Materials.
At convergence of Algorithm~\ref{algoMFVB}, inference focuses directly on the marginal approximate posterior for the regression coefficients  $q\mf^\star(\bbeta) = \phi_p(\bbeta - \bmu\mf^\star,\bSigma\mf^\star)$.

\begin{algorithm*}[t]
\caption{Mean-Field Variational Bayes for the cumulative probit model}
	\label{algoMFVB}
	\vspace{3pt}	
	{[{\bf 1}]} {Precompute $\bV=(\bSigma_0^{-1} + \bX\T\bX)^{-1}$ and $\bSigma_0^{-1}\bmu_0$ and initialize $\bar{\bz} = (\bar{z}_1, \dots, \bar{z}_n)\T$}\\
	\vspace{1pt}
	{[{\bf 2}]}      \While(){increases in $\log \papp(\by,q\mf)$ are larger than a small value $\varepsilon$}
	{
		\vspace{1pt}
		Set $\bar{\bbeta} = \bV (\bSigma_0^{-1}\bmu_0 + \bX \bar{\bz} )$ \\
		\vspace{1pt}
		Set $\bar{z}_i = \bxit\bar{\bbeta} - \zeta_1(\alpha_{y_i-1} -\bxit\bar{\bbeta}, \alpha_{y_i} -\bxit\bar{\bbeta})$ for each $i=1,\dots,n$\\
}
	\vspace{1pt}
	{[{\bf 3}]} {Set $\bmu\mf^\star = \bar{\bbeta}$ and $\bSigma\mf^\star = \bV$} \\
	\vspace{1pt}
	{\bf Output:} {\small Optimal \textsc{mfvb} approximation $q\mf^\star(\bbeta) = \phi_p(\bbeta - \bmu\mf^\star, \bSigma\mf^\star)$}
\end{algorithm*}

\subsection{Partially factorized Mean-Field}
\label{sec:pmf}
The partially factorized Mean-Field (\textsc{pmf}) approach developed in \citet{fasano:2022} focuses on improving standard \textsc{mfvb} by introducing a larger variational family $\mathcal{Q}\pmf$ that includes $\mathcal{Q}\mf$ as a special case.
A primary step for developing such a routine in the cumulative probit is to generalize the approach of \citet{holmes:2006} to the ordinal case, noticing that from Equation~\eqref{eq:full_cond} the posterior distribution $p(\bbeta,\bz\mid\by)$ factorizes as the product between $p(\bbeta\mid\bz)$ and $p(\bz\mid\by)$, defined as
\begin{equs}[e:block]
\label{eq:pmf_fact}
p(\bbeta\mid\bz) &= \phi_p(\bbeta - \bV(\bSigma_0^{-1}\bmu_0 + \bX\T\bz),\bV),\\
p(\bz\mid\by) &\propto \phi_n(\bz -\bX\bmu_0, \mathbf{I}_n + \bX\bSigma_0\bX\T) \prod_{i=1}^n\mathds{I}[\alpha_{y_i-1} < z_i < \alpha_{y_i}].
\end{equs}
The \textsc{pmf} family is specified as $\mathcal{Q}\pmf= \left\{q\pmf(\bbeta,\bz): q\pmf(\bbeta,\bz) = q\pmf(\bbeta\mid\bz)\prod_{i=1}^n q\pmf(z_i) \right\}$, and according to Theorem~2 of \citet{fasano:2022} the minimization of the \textsc{kl} divergence to find the optimal densities in $\mathcal{Q}\pmf$ leverages the decomposition
\begin{equs}[e:block]
	\label{eq:kl_pmf}
 {\textsc{kl}}\left[q\pmf(\bbeta, \bz) \,||\, p(\bbeta,\bz\mid\by) \right]  = {\textsc{kl}}\left[q\pmf(\bz) \,||\, p(\bz\mid\by) \right]  + \mathbb{E}_{q\pmf(\bz)} \big\{{\textsc{kl}}\left[q\pmf(\bbeta\mid\bz) \,||\, p(\bbeta\mid\bz) \right]\big\}.\quad
\end{equs}
Notably, Equation~\eqref{eq:kl_pmf} implies that the second term is zero if and only if the optimal density $q^\star\pmf(\bbeta\mid\bz) = p(\bbeta\mid\bz)$ as in~\eqref{eq:pmf_fact}, and according to such a factorization the optimal densities for the latent variables $z_i$ in $\mathcal{Q}\pmf$ satisfy $q\pmf^\star(z_i) \propto \exp\left\{\mathbb{E}_{q^\star\pmf(z\mi)} \log p(z_i \mid \bz\mi, \by)\right\}$ for each $i=1,\dots,n$, where $\bz\mi$ denotes the vector $\bz$ without the element $z_i$.
Such conditional distributions can be obtained following the derivations provided in Section~\ref{supp:pmf} of the Supplementary Materials as
\begin{equs}
	(z_i\mid \bz \mi,\by) \sim \textsc{tn}\left([\alpha_{y_i-1}, \alpha_{y_i}],
 \bxit\bmu_0 + \frac{1}{1-\bxit\bV\bxii} \bxit \bV \bX\mi\T \left(\bz\mi - \bX\mi\bmu_0\right), \frac{1}{1-\bxit\bV\bxii}\right);
\end{equs}
therefore, the optimal densities $q\pmf^\star(z_i)$ are also Truncated-Normals over the interval $[\alpha_{y_i-1}, \alpha_{y_i}]$ with optimal scales $\sigma^\star_i = (1-\bxit\bV\bxii)^{-1/2}$ and location parameters $\xi_i$ that can be iteratively updated via variational expectations until convergence.

\begin{algorithm*}[t]
\caption{Partially factorized Mean-Field for the cumulative probit model}
	\label{algoPMFVB}
	\vspace{3pt}	
	{[{\bf 1}]} {Precompute $\bV=(\bSigma_0^{-1} + \bX\T\bX)^{-1}$, $\sigma^\star_i = (1-\bxit\bV\bxii)^{-1/2}$ and initialize $\bar{\bz} = (\bar{z}_1, \dots, \bar{z}_n)\T$}\\
	\vspace{1pt}
	{[{\bf 2}]}      \While(){increases in $\log \papp(\by,q\pmf)$ are larger than a small value $\varepsilon$}
	{

		\For(){$i=1,\dots,n$}
		{
		\vspace{1pt}
		Set $\xi_i = \bxit\bmu_0 + {\sigma_i^\star}^2 \bxit \bV \bX\mi\T \left(\bar{\bz}\mi - \bX\mi\bmu_0\right)$\\
		Set $\bar{z}_i = \xi_i - \sigma_i^\star \zeta_1(\tilde{u}_i, \tilde{v}_i)$ with $\tilde{u}_i = (\alpha_{y_i-1} - \xi_i)/\sigma_i^\star$ and  $\tilde{v}_i = (\alpha_{y_i} - \xi_i)/\sigma_i^\star$
}}
	\vspace{1pt}
	{[{\bf 3}]} {Set $\xi^\star_i = \xi_i$,  $\tilde{u}_i^\star = (\alpha_{y_i-1} - \xi^\star_i)/\sigma_i^\star$ and  $\tilde{v}_i^\star = (\alpha_{y_i} - \xi_i^\star)/\sigma_i^\star$ for $i=1,\ldots,n$}\\
	{[{\bf 4}]}  {{\em (optional)} Set $\bar{\bz}^\star=(\bar{z}_1^\star,\ldots,\bar{z}_n^\star)\T$ and ${\bOmega}^\star = \mbox{diag}(\omega^\star_1,\ldots, \omega^\star_n)$ with $\bar{z}_i^\star = \xi_i^\star - \sigma_i^\star \zeta_1(\tilde{u}_i^\star, \tilde{v}_i^\star)$\\
	\hphantom{[{\bf 4}] {\em (optional)}}  and $\omega^\star_i = {\sigma^\star_i}^2(1-\zeta_1(\tilde{u}_i^\star, \tilde{v}_i^\star)^2 - \zeta_2(\tilde{u}_i^\star, \tilde{v}_i^\star))$ for $i=1,\ldots,n$}\\
	\vspace{1pt}
	{\bf Output:} {\small Optimal \textsc{pmf} density $q\pmf^\star(z_i) = [\Phi(\tilde{v}_i^\star) - \Phi(\tilde{u}_i^\star)]^{-1}\phi(z_i - \xi^\star_i, {\sigma_i^\star}^2)\mathds{I}[\alpha_{y_i-1} < z_i < \alpha_{y_i}]$\\
	{\em(optional)}\, Optimal $q\pmf^\star(\bbeta)$ moments $\bmu\pmf^\star =\bV(\bSigma_0^{-1}\bmu_0 + \bX\T \bar{\bz}^\star)$, $\bSigma^\star\pmf=\bV + \bV\bX\T{\bOmega}^\star \bX\bV$}
\end{algorithm*}

Algorithm~\ref{algoPMFVB} illustrates an iterative procedure for obtaining such parameters and, optionally, the marginal moments $\bmu\pmf^\star = \mathbb{E}_{q\pmf^\star(\bbeta)}[\bbeta]$ and $\bSigma\pmf^\star=\mbox{var}_{q\pmf^\star(\bbeta)}[\bbeta]$ relying on the intermediate functions $\zeta_1(\cdot,\cdot)$ and $\zeta_2(a,b) = [b\phi(b) - a\phi(a)][\Phi(b) - \Phi(a)]^{-1}$ for computing univariate Truncated-Normal moments.
Indeed, according to the factorization of $\mathcal{Q}\pmf$, the optimal joint density for the local variables is obtained as $q^\star\pmf(\bz) = \prod_{i=1}^n q^\star\pmf(z_i)$, and  the marginal $q^\star\pmf(\bbeta)$ is obtained as
\begin{equs}[e:block]
	\label{eq:pmf_beta_marg}
q^\star\pmf(\bbeta)= \int q^\star\pmf(\bbeta\mid\bz)q^\star\pmf(\bz)\mbox{d}\bz = \int p(\bbeta\mid\bz)\prod_{i=1}^n q^\star\pmf(z_i)\mbox{d}\bz.
\end{equs}
In Algorithm~\ref{algoPMFVB}, the marginal moments are obtained from~\eqref{eq:pmf_beta_marg} via iterated expectations as
\begin{equs}
	\label{eq:pmf_moments}
&\bmu\pmf^\star = \mathbb{E}_{q\pmf^\star(\bbeta)}[\bbeta] = \mathbb{E}_{q\pmf^\star(\bz)}\left[\mathbb{E}_{p(\bbeta\mid\bz)}[\bbeta]\right] =\bV\left(\bSigma_0^{-1}\bmu_0 + \bX\T \mathbb{E}_{q\pmf^\star(\bz)}\left[\bz\right]\right)\\
&\bSigma\pmf^\star = \mbox{var}_{q\pmf^\star(\bbeta)}[\bbeta] = \mathbb{E}_{q\pmf^\star(\bz)}\left[\mbox{var}_{p(\bbeta\mid\bz)}[\bbeta]\right] + \mbox{var}_{q\pmf^\star(\bz)}\left[\mathbb{E}_{p(\bbeta\mid\bz)}[\bbeta]\right]  = \bV + \bV\bX\T\left\{\mbox{var}_{q\pmf^\star(\bz)}[\bz]\right\} \bX\bV.
\end{equs}
At the end of each iteration, the \textsc{elbo} can be computed analytically to monitor convergence as
\begin{equs}[e:block]
\label{eq:elbo_pmf}
\log\papp(\by,q\pmf) &= 	\sum_{i=1}^n \log [\Phi(\tilde{v}_i) - \Phi(\tilde{u}_i)] + \frac{1}{2}\sum_{i=1}^n\frac{ \bar{z}_i^2}{{\sigma_i^\star}^2} - \sumin\frac{\bar{z}_i\xi_i}{{\sigma_i^\star}^2} + \frac{1}{2}\sumin\frac{\xi_i^2}{{\sigma_i^\star}^2}\\
		 & - \frac{1}{2} \bar{\bz}\T(\mathbf{I}_n - \bX\bV\bX\T) \bar{\bz} + \bar{\bz}\T(\mathbf{I}_n - \bX\bV\bX\T)\bX\bmu_0 + const,
\end{equs}
where ``${const}$'' denotes an additive constant not depending on the variational parameters;
refer to Section~\ref{supp:pmf_elbo} in the Supplementary Materials the complete derivation, and to Section~\ref{supp:pmf_comp} for further computational considerations.

In most settings, the moments $\bmu\pmf^\star$ and $\bSigma\pmf^\star$ are often the main focus of empirical analysis; however, the \textsc{pmf} routine easily allows full posterior inference on any functional of the marginal approximate density $q^\star\pmf(\bbeta)$ via Monte Carlo integration, generalizing Proposition 2 of \citet{fasano:2022}.
Indeed, a realization from the approximate density for $\bbeta$ is obtained by sampling each component $z_i$ of $\bz$ independently from its optimal Truncated-Normal restricted to the interval $[\alpha_{y_{i}-1}, \alpha_{y_i}]$ with location $\xi_i^\star$ and scale $\sigma_i^\star$ for each $i=1,\ldots,n$, and using such a realization to sample $\bbeta$ from its conditional Gaussian $p(\bbeta\mid\bz)$.

\subsection{Expectation Propagation}
\label{sec:ep}

The overarching goal of Expectation Propagation (\textsc{ep}) is to approximate the posterior distribution $p(\bbeta\mid\by)$ with a more manageable density $q\ep(\bbeta)$, frequently assumed to belong to the Gaussian family.
This quantity is constructed avoiding augmentation schemes with local variables, and directly to mimic the product structure of the posterior distribution~\eqref{eq:posterior}, normalizing the product of $(n+1)$ tractable factors (or {\em sites}) $q_i$ associated to the likelihood terms $\ell_i$ and the prior $p_0$ as
\begin{equs}[e:block]
	\label{eq:q_ep}
	q\ep(\bbeta) = \frac{1}{Z\ep} q_0(\bbeta)\prod_{i=1}^n q_i(\bbeta)=\frac{1}{Z\ep} \prod_{i=0}^n q_i(\bbeta).
\end{equs}
Assuming that each site is a Gaussian density function for $\bbeta$ with natural parameters $(\bQ_i, \br_i)$ and normalizing constant $Z_i$ for $i=0,\ldots,n$,
\begin{equs}[e:block]
	\label{eq:q_iep}
	q_i(\bbeta) = \frac{1}{Z_i}  \exp\left\{-\frac{1}{2} \bbeta\T \bQ_i \bbeta + \bbeta\T \br_i \right\}
	\quad\mbox{and}\quad q\ep(\bbeta)= \frac{1}{Z\ep}\exp\left\{-\frac{1}{2} \bbeta\T  \bQ\ep \bbeta + \bbeta\T \br\ep \right\},
\end{equs}
with $\bQ\ep = \sum_{i=0}^n \bQ_i$ and $\br\ep = \sum_{i=0}^n \br_i$; therefore, the global \textsc{ep} approximation is again Gaussian
$q\ep(\bbeta) = \phi_p(\bbeta - \bmu\ep, \bSigma\ep)$ with $\bSigma\ep = \bQ\ep^{-1}$ and $\bmu\ep = \bQ\ep^{-1}\br\ep$.

The ideal goal of \textsc{ep} is to determine the optimal parameters of $q\ep(\bbeta)$ by minimizing the {\em reverse} \textsc{kl} divergence ${\textsc{kl}}\left[p(\bbeta \mid \by) \,\|\, q\ep(\bbeta)\right]$, which, compared to~\eqref{eq:klvb}, reverses the arguments and direction \citep{minkaEP}.
Since $q\ep(\bbeta)$ belongs to the exponential family, the minimum is obtained when the moments of its sufficient statistics computed with respect to $q\ep(\bbeta)$ and $p(\bbeta \mid \by)$ coincide  \citep[e.g.,][Section 10.7]{bishop:2006}; however, this approach is impractical as it requires evaluating expectations with respect to the intractable posterior distribution $p(\bbeta \mid \by)$.
The core idea of \textsc{ep}, conceptually and algorithmically, is to iteratively optimize each site $q_i(\bbeta)$ in~\eqref{eq:q_ep} by minimizing a tractable approximation to the reverse \textsc{kl} divergence where, at each iteration, the exact posterior is approximated by the product between the likelihood term associated with the site being updated and the posterior approximation excluding that site \citep{opper:2000,minkaEP}.

To clarify this aspect, the update of a generic site $q_i(\bbeta)$---while keeping all other site approximations fixed at their current values---relies on the intermediate distributions
\begin{equs}[e:block]
\label{eq:cavity_hybrid}
q\mi(\bbeta) \propto \frac{q\ep(\bbeta)}{q_i(\bbeta)} &= \frac{1}{Z\mi} \exp\left\{-\frac{1}{2} \bbeta\T \bQ\mi \bbeta + \bbeta\T \br\mi \right\}
\quad\mbox{and}\quad h_i(\bbeta) &= \frac{1}{Z_{h_i}} \ell_i(\bbeta) q\mi(\bbeta),
\end{equs}
commonly referred to as the {\em cavity} and {\em hybrid} distributions, and corresponding to the posterior approximation that excludes the $i$-th site and its product with the $i$-th likelihood contribution, respectively; according to~\eqref{eq:q_iep}, and leveraging the exponential family specification for $q_i(\bbeta)$, the natural parameters of the cavity are $\bQ\mi=\sum_{l\neq i}\bQ_l$, $\br\mi=\sum_{l\neq i}\br_l$.
The key of \textsc{ep} procedures is to update the quantities $(Z_i, \bQ_i, \br_i)$ that parametrize $q_i(\bbeta)$ matching the moments of the hybrid $h_i(\bbeta)$ with those of the resulting global approximation, obtained as the product between the cavity and $q_i(\bbeta)$ itself; denoting as $\bmu_{h_i} := \mathbb{E}_{h_i(\bbeta)}[\bbeta]$, $\bSigma_{h_i} := \mbox{var}_{h_i(\bbeta)}[\bbeta]$ and following \citet{anceschi:2024EP}, these conditions require
\begin{equs}[e:block]
	\label{eq:moment_cond}
	\bSigma_{h_i}=	(\Qmi + \bQ_i)^{-1}, \quad \bmu_{h_i} = (\Qmi + \bQ_i)^{-1}(\rmi + \br_i), \quad Z_{h_i} = \frac{\Psi(\Qmi+ \bQ_i, \rmi + \br_i)}{Z\mi Z_i};
\end{equs}
solving for $(Z_i, \bQ_i, \br_i)$ the moment conditions can be imposed as
\begin{equs}[e:block]
	\label{eq:moment_sol}
	\bQ_i &= \bSigma_{h_i}^{-1} - \Qmi, \quad \br_i &= \bSigma_{h_i}^{-1}\bmu_{h_i} - \rmi,\\
	\multicol{4}{\log Z_i  = \log \Psi(\Qmi + \bQ_i , \br\mi + \br_i) -Z\mi - \log Z_{h_i}},
\end{equs}
where $\Psi(\bQ,\br)$ denotes the normalizing constant of a Gaussian with natural parameters $(\bQ,\br)$,
\begin{equs}
	\Psi(\bQ,\br) &= \int \exp\left\{-\frac{1}{2} \bbeta\T \bQ \bbeta + \bbeta\T \br \right\} \mbox{d}\bbeta
		      &= \frac{(2\pi)^{p/2}}{|\bQ|^{1/2}}\exp\left\{\frac{1}{2}\br\T\bQ^{-1}\br\right\}.
\end{equs}

The procedure is iterated sequentially across all sites  $q_i(\bbeta)$ $i=1,\dots,n$ until convergence; since the prior $p_0$ is Gaussian, $q_0$ is directly matched with the prior letting $\bQ_0 = \bSigma_0^{-1}$ and $\br_0 = \bSigma_0^{-1}\bmu_0$ \citep{pima,anceschi:2024EP}.
Notably, the quantities $(Z_{h_i}, \bSigma_{h_i}, \bmu_{h_i})$ required in~\eqref{eq:moment_sol} can be derived avoiding numerical integration or complex algebraic derivations that commonly characterize \textsc{ep} procedures \citep[e.g.,][]{kim:2016,zhou:2023}; indeed, under the cumulative probit model with likelihood~\eqref{eq:likelihood}, the hybrid distribution $h_i(\bbeta)$
\begin{equs}[e:block]
\label{eq:hybrid}
h_i(\bbeta) &= \frac{1}{Z_{h_i}}\ell_i(\bbeta)  q_{-i}(\bbeta)  \\
	    &= \frac{1}{Z_{h_i}}\left[\Phi(\alpha_{y_i} -\bxit\bbeta) - \Phi(\alpha_{y_i-1} -\bxit\bbeta)\right]\phi_p(\bbeta-\bQ^{-1}\mi\br\mi, \bQ^{-1}\mi)
\end{equs}
is recognizable as the kernel of a Selection-Normal distribution \citep{arellano:2006}.
Following the details provided in Section~\ref{appendix:selnorm} of the Supplementary Materials, the required moments are obtained letting
\begin{equs}[e:block]
\label{eq:lim_z}
\bar{u}_i = \frac{\alpha_{y_i -1} - \bx_i^\intercal\Qmi^{-1}\rmi}{(1+\bx_i^\intercal\Qmi^{-1}\bx_i)^{1/2}}, \quad \bar{v}_i = \frac{\alpha_{y_i} - \bx_i^\intercal\Qmi^{-1}\rmi}{(1+\bx_i^\intercal\Qmi^{-1}\bx_i)^{1/2}},
\end{equs}
and computing analytically $Z_{h_i} = \Phi(\bar{v}_i) - \Phi(\bar{u}_i)$ and
\begin{equs}[e:block]
	\label{eq:selN_moments}
	\bmu_{h_i} &= \Qmi^{-1} \br\mi -  \frac{\zeta_{1}(\bar{u}_i,\bar{v}_i )}{(1+\bx_i^\intercal\Qmi^{-1}\bx_i)^{1/2} } \Qmi^{-1}\bx_i\\
\bSigma_{h_i} &= \Qmi^{-1} - \left(\frac{\zeta_{1}(\bar{u}_i,\bar{v}_i)^2 + \zeta_{2}(\bar{u}_i,\bar{v}_i)}{1+\bx_i^\intercal\Qmi^{-1}\bx_i}\right) \Qmi^{-1}\bx_i\bx_i^\intercal\Qmi^{-1}.
\end{equs}
These derivations effectively generalize \textsc{ep} routines for the binary probit, as developed in  Section 3.6 of \citet{gp} for Gaussian processes and Section 4.1 of \citet[]{anceschi:2024EP} for a regression setting; see also Section 3.2 of \citet{chu:2005}.
According to~\eqref{eq:selN_moments}, the moments of the hybrid and the cavity are related via simple rank-one operations, and this relation translates also to the updates of the global \textsc{ep} parameters \citep{anceschi:2024EP}. Algorithm~\ref{algoEP} illustrates the iterative updates that characterize the \textsc{ep} routine, more conveniently parametrized in terms of covariance matrices $\Sep = \Qep^{-1}$ and $\Smi=\Qmi^{-1}$; refer to Section~\ref{appendix:ep} in the Supplementary Materials for the complete derivation.
Convergence of the \textsc{ep} routine can be assessed by monitoring a suitable approximation of the marginal likelihood, obtained replacing the likelihood terms with their site approximations (e.g., \citealp{vehtari:2020}, Appendix E; \citealp{anceschi:2024EP}, Section 2.2) and letting
\begin{equs}[e:block]
\label{eq:papp_ep}
\log \papp(\by,q\ep) := \log \int  q_{0}(\bbeta) \prod_{i=1}^n q_i(\bbeta)\mbox{d}\bbeta = \log \Psi(\rep,\Qep) - \log \Psi(\br_0, \bQ_0) - \sum_{i=1}^n \log Z_i.
\end{equs}

\begin{algorithm*}[t]
\caption{Expectation Propagation for the cumulative probit model}
	\label{algoEP}
	\vspace{3pt}	
	{[{\bf 1}]} {Initialize $\Sep = \bSigma_0$, $\br\ep = \bSigma_0^{-1}\bmu_0$ and the scalars $(k_i,w_i)$ for $i=1, \dots, n$}\\
	\vspace{1pt}
	{[{\bf 2}]}      \While(){changes in $\log \papp(\by,q\ep)$ are larger than a small value $\varepsilon$}
	{
		\For(){$i=1,\dots,n$}
		{
		\vspace{1pt}
		Set $\Smi = \Sep + [1-k_i \bx_i\T\Sep\bx_i]^{-1}k_i \Sep\bx_i\bx_i\T \Sep$, $\rmi = \rep  - w_i \bx_i$ and
		${\bar{u}_i = (\alpha_{y_i -1} - \bx_i\T\Smi\rmi){\left[1+\bx_i\T\Smi\bx_i\right]^{-1/2}}},{\bar{v}_i = (\alpha_{y_i} - \bx_i\T\Smi\rmi){\left[1+\bx_i\T\Smi\bx_i\right]^{-1/2}}}$\\[2pt]
		Set $k_i = {\left[\vphantom{\Smi}\zeta_{1}(\bar{u}_i,\bar{v}_i)^2 + \zeta_{2}(\bar{u}_i,\bar{v}_i)\right]}{\left[1 + \bx_i\T \Smi \bx_i [1 - {\zeta_{1}(\bar{u}_i,\bar{v}_i)^2 - \zeta_{2}(\bar{u}_i,\bar{v}_i)}]\right]^{-1}}$ and update
		$\Sep = \Smi - \left[\zeta_{1}(\bar{u}_i,\bar{v}_i)^2 + \zeta_{2}(\bar{u}_i,\bar{v}_i)\right]\left[1+\bx_i\T\Smi\bx_i\right]^{-1} \Smi\bx_i\bx_i\T\Smi$\\[2pt]
		Set $w_i = (\bx_i\T \Smi \rmi)k_i - \zeta_1(\bar{u}_i, \bar{v}_i) (1+\bxit\Smi\bx_i)^{1/2}$ and update $\rep = \rmi  + w_i \bx_i$\\[2pt]
	Set $Z_i = (1+k_i \bxit\Smi\bxii)^{-1}\left[2 w_i \bxit\Smi\rmi + w_i^2 \bxit \Smi\bxii - k_i (\rmi\T\Smi\bxii)^2\right]$
}
}
	\vspace{1pt}
	{[{\bf 3}]} {Set $\bmu\ep^\star = \Sep\br\ep$ and $\bSigma\ep^\star = \Sep$} \\
	\vspace{1pt}
	{\bf Output:} {\small Optimal \textsc{ep} approximation $q\ep^\star(\bbeta) = \phi_p(\bbeta - \bmu\ep^\star, \bSigma\ep^\star)$}
\end{algorithm*}

\subsection{Estimation of the thresholds and prediction}
\label{sec:alphaopt}
The proposed \textsc{mfvb}, \textsc{pmf} and \textsc{ep} algorithms focus on providing estimates of the posterior distribution for the regression coefficients under fixed threshold values $\balpha$; in practice, however, these quantities are unknown and must be estimated from the data. 
We propose an Empirical Bayes approach to estimation, optimizing the marginal likelihood of the cumulative probit with respect to such quantities.
As discussed in the previous Sections, each routine induces a tractable approximation of the marginal likelihood $p(\by)$, denoted as $\papp(\by, q)$ with $q\in\{q\mf, q\pmf, q\ep\}$ depending on the specific method; following a common practice for variational approximations of mixed models in frequentist inference \citep[e.g.,][]{ormerod:2012,hall2020fast,gronberg:2025}, maximization of the approximate marginal likelihood  can be carried out using derivative-free methods such as those implemented in the \texttt{R} function \texttt{optim}; to facilitate optimization, it is convenient to reparameterize the thresholds into log-increments  $\tau_k = \log(\alpha_{k} - \alpha_{k-1})$ for ${k = 2,\dots,K-1}$ and $\tau_1 = \alpha_1$.
Since $K$ is typically small in practical applications (e.g., $K = 5$ for standard Likert scales), the procedure remains computationally feasible, provided that the evaluation of the approximate marginal likelihood is sufficiently fast.

A more efficient optimization approach relies on computing the gradient of the approximate marginal likelihood with respect to $\balpha$ analytically; this can be done efficiently under the \textsc{mfvb} routine, since from~\eqref{eq:elbo_full}
\begin{equs}[e:block]
\label{eq:alpha_deriv}
\frac{\partial\, \papp(\by,q\mf)}{\partial \, \alpha_k} = \sum_{i=1}^n \frac{\mathds{I}[y_i = k]\cdot \phi(\alpha_{y_i} -\bxit\bar{\bbeta}) - \mathds{I}[(y_i-1) = k]\cdot \phi(\alpha_{y_i - 1} -\bxit\bar{\bbeta})}{\Phi(\alpha_{y_i} -\bxit\bar{\bbeta}) - \Phi(\alpha_{y_i - 1} -\bxit\bar{\bbeta})}.
\end{equs}
At convergence of the \textsc{mfvb} routine, $\bar{\bbeta}$ is treated as fixed quantity, and its dependence on $\balpha$ is ignored; $\papp(\by, q\mf)$ is therefore optimized with respect to $\balpha$ by setting \eqref{eq:alpha_deriv} to zero and applying Newton’s method.
This maximization for a given $\bar{\bbeta}$ can be efficiently performed using existing software for maximum likelihood estimation of the cumulative probit model \citep[e.g.,][]{ordinal}, treating $\bxit\bar{\bbeta}$ as an offset term.
The updated $\balpha$ is then provided as a new value to the main \textsc{mfvb} routine to obtain a new estimate of $\bar{\bbeta}$, and this alternating process is repeated until convergence, which generally occurs in a few iterations.

Under the \textsc{pmf} routine, the derivatives of the \textsc{elbo} in \eqref{eq:elbo_pmf} with respect to $\balpha$ are more involved and no longer coincide with~\eqref{eq:alpha_deriv}; under the \textsc{ep} approximation, $\papp(\by, q\ep)$ does not admit a closed form expression and its gradient with respect to $\balpha$ cannot be computed analytically. 
Nevertheless, we have found that employing the same alternating optimization strategy based on equating \eqref{eq:alpha_deriv} to zero with $\bar{\bbeta} = \bmu\pmf^\star$ or $\bar{\bbeta} = \bmu\ep^\star$ provides a reasonable search direction that dramatically improves over derivative free methods, and yields good empirical performance.

Predictions for future observations can be computed by leveraging such estimated thresholds, denoted as ${(-\infty,\hat{\alpha}_1, \ldots,\hat{\alpha}_{K-1},\infty)}$, and the resulting approximations for the regression coefficients.
Under \textsc{mfvb} and \textsc{ep}, the approximate posterior is Gaussian and the predictive probabilities for a new statistical unit with covariates $\bx\new \in \mathbb{R}^p$ can be computed in closed form; focusing on $q\ep^\star(\bbeta) = \phi_p(\bbeta - \bmu\ep^\star, \bSigma\ep^\star)$ and using Lemma 7.1 of \citet{azza},
\begin{equs}
	\mbox{pr}({y}\new \leq k\mid\by) &= \mathbb{E}_{q\ep^\star(\bbeta)}[\Phi(\hat{\alpha}_{k} - \bx\new^\intercal\bbeta)]
					 &= \Phi\left(\frac{\hat{\alpha}_{k} -\bx\new^\intercal\bmu\ep^\star}{(1+\bx\new^\intercal\bSigma^\star\ep\bx\new)^{1/2}}\right).
\end{equs}
Similarly, class-specific predictive probabilities are obtained as
\begin{equs}
	\mbox{pr}({y}\new = k\mid\by) &= 
	\mathbb{E}_{q\ep^\star(\bbeta)}[\Phi(\hat{\alpha}_{k} - \bx\new^\intercal\bbeta) - \Phi(\hat{\alpha}_{k-1} -  \bx\new^\intercal\bbeta)] \\
					 &=\Phi\left(\frac{\hat{\alpha}_{k} -\bx\new^\intercal\bmu\ep^\star}{(1+\bx\new^\intercal\bSigma\ep^\star\bx\new)^{1/2}}\right) - \Phi\left(\frac{\hat{\alpha}_{k-1} -\bx\new^\intercal\bmu\ep^\star}{(1+\bx\new^\intercal\bSigma\ep^\star\bx\new)^{1/2}}\right)
\end{equs}
and future observations can be assigned to the category that maximizes such predictive probabilities.
Under the \textsc{pmf} routine, predictive probabilities are not available in closed form since 
\begin{equs}
	\mathbb{E}_{q\pmf^\star(\bbeta)}[\Phi(\hat{\alpha}_{k} - \bx\new^\intercal\bbeta)] &=	\mathbb{E}_{q\pmf^\star(\bz)}\left[\mathbb{E}_{p(\bbeta\mid \bz)}[\Phi(\hat{\alpha}_{k} - \bx\new^\intercal\bbeta)]\right]\\ &= \mathbb{E}_{q\pmf^\star(\bz)}\left[\Phi\left(\frac{\hat{\alpha}_{k} -\bx\new\T\bV(\bSigma_0^{-1}\bmu_0 + \bX\T\bz)}{(1+\bx\new^\T\bV\bx\new)^{1/2}}\right)\right]
\end{equs}
involves non-linear functionals of $q\pmf^\star(\bz)$; as outlined in Section~\ref{sec:pmf}, it is straightforward to sample from the tractable $q\pmf^\star(\bz)$ with independent Truncated-Normal components and evaluate such predictive probabilities via Monte Carlo integration; see also Corollary 1 of \citet{fasano:2022}.

\section{Simulation studies}
\label{sec:sim}

The empirical properties of the proposed approximations are evaluated in a simulation study, aimed at assessing their accuracy and computational efficiency under different scenarios.
We focus on settings with large sample size and modest number of covariates, and consider varying sample sizes $n \in \{500, 1000, 2500, 5000, 10000\}$ and numbers of covariates $p \in \{5, 25, 50\}$, fixing the true coefficient vector such that $20\%$ of its elements are equal to $0$, $40\%$ to $1$, and $40\%$ to $-1$. 
Covariates are independently drawn from a uniform distribution on $[0,1]$, and each column of the resulting design matrix is rescaled to have mean $0$ and standard deviation $0.5$. The ordinal response variable is then generated in two steps following Equation~\ref{eq:mod_aug}, simulating for each observation $i = 1, \dots, n$ a Gaussian latent response with unit variance and mean $\bxit\bbeta$, and discretizing the resulting values into $K = 5$ ordinal categories, using ordered cutoffs sampled uniformly over the observed range of values.

\begin{figure}[tb]
\centering
\includegraphics[width=\textwidth]{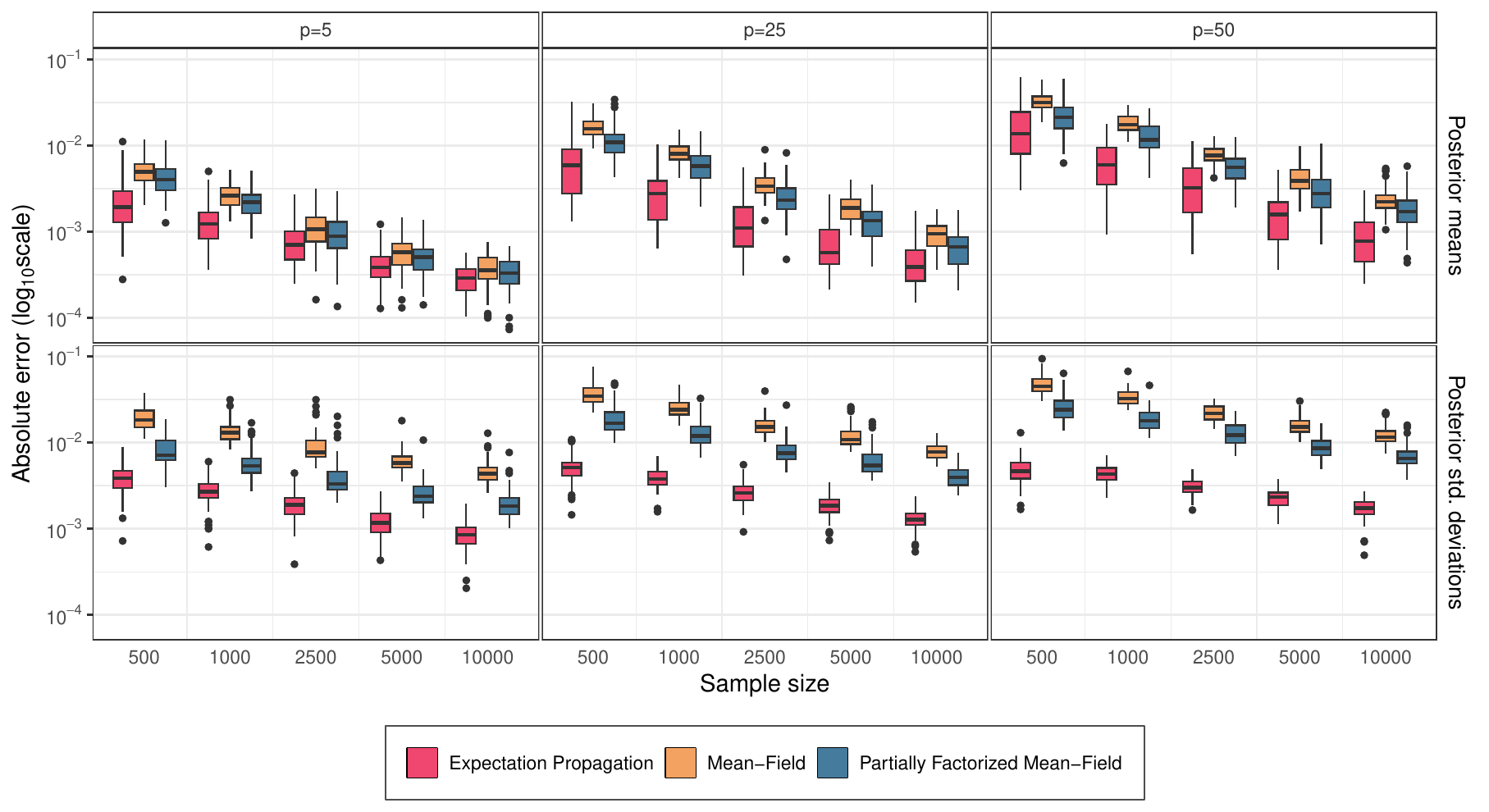}
\caption{Average absolute differences between posterior estimates obtained via \textsc{mcmc} and the proposed approximations. Values are displayed on log-scales for graphical clarity; boxplots represent variability across $100$ simulation replications.}
\label{fig:sim}
\end{figure}

The first objective of the simulation studies is to compare the ability of the three proposed methods to estimate the marginal posterior means and standard deviations of the ``true'' posterior for the regression coefficients,  as obtained from \textsc{mcmc} samples.
In all settings, the regression coefficients are assigned a Gaussian prior with mean zero and diagonal covariance with elements equal to $2$; the proposed \textsc{mfvb}, \textsc{pmf} and \textsc{ep} procedures are implemented using a convergence criterion of $\varepsilon = 10^{-6}$ and optimizing the cutoffs as described in Section~\ref{sec:alphaopt}.
Posterior sampling is implemented in \textsc{stan} \citep{rstan}, considering a full Bayesian model specification that treats the thresholds as additional parameters with a uniform prior over the unconstrained log-increments. 
Sampling is performed for $5000$ iterations following a burn-in of $1000$, and for each method that provides an approximate posterior $q^\star$, the quality of the estimated posterior means and standard deviations is measured as
\begin{equs}
	\frac{1}{p} \sum_{j=1}^p \big|\mathbb{E}_{q^\star(\beta_j)}[\beta_j]-\mathbb{E}_{p(\beta_j\mid\by)}[\beta_j]\big| \quad\mbox{and}\quad
\frac{1}{p} \sum_{j=1}^p\Big|\big(\mbox{var}_{q^\star(\beta_j)}[\beta_j]\big)^{0.5}-\big(\mbox{var}_{p(\beta_j\mid\by)}[\beta_j]\big)^{0.5}\Big|.
\end{equs}
Figure~\ref{fig:sim} provides a graphical comparison of these quantities, based on $100$ independent replications for each scenario to assess simulation variability. 
Current empirical findings indicate strong performance for all methods, with accuracy improving as the sample size increases and with a smaller number of covariates.
For instance, in the setting with $n = 10000$ and $p = 5$, the error in the estimated posterior means is below $10^{-3}$ for all methods across all replications, while for $p = 50$ it remains below $10^{-2}$. 
Notably, \textsc{ep} yields the most accurate approximation of the posterior means and performs particularly well in capturing posterior uncertainty, providing the best approximations of the posterior standard deviations across all settings. 
These findings are consistent with previous empirical evidence \citep[e.g.,][]{pima,vehtari:2020,anceschi:2023,zhou:2023}, which attributes the superior performance of \textsc{ep} to its iterative refinement around regions of high posterior density.

\begin{figure}[tb]
\centering
\includegraphics[width=\textwidth]{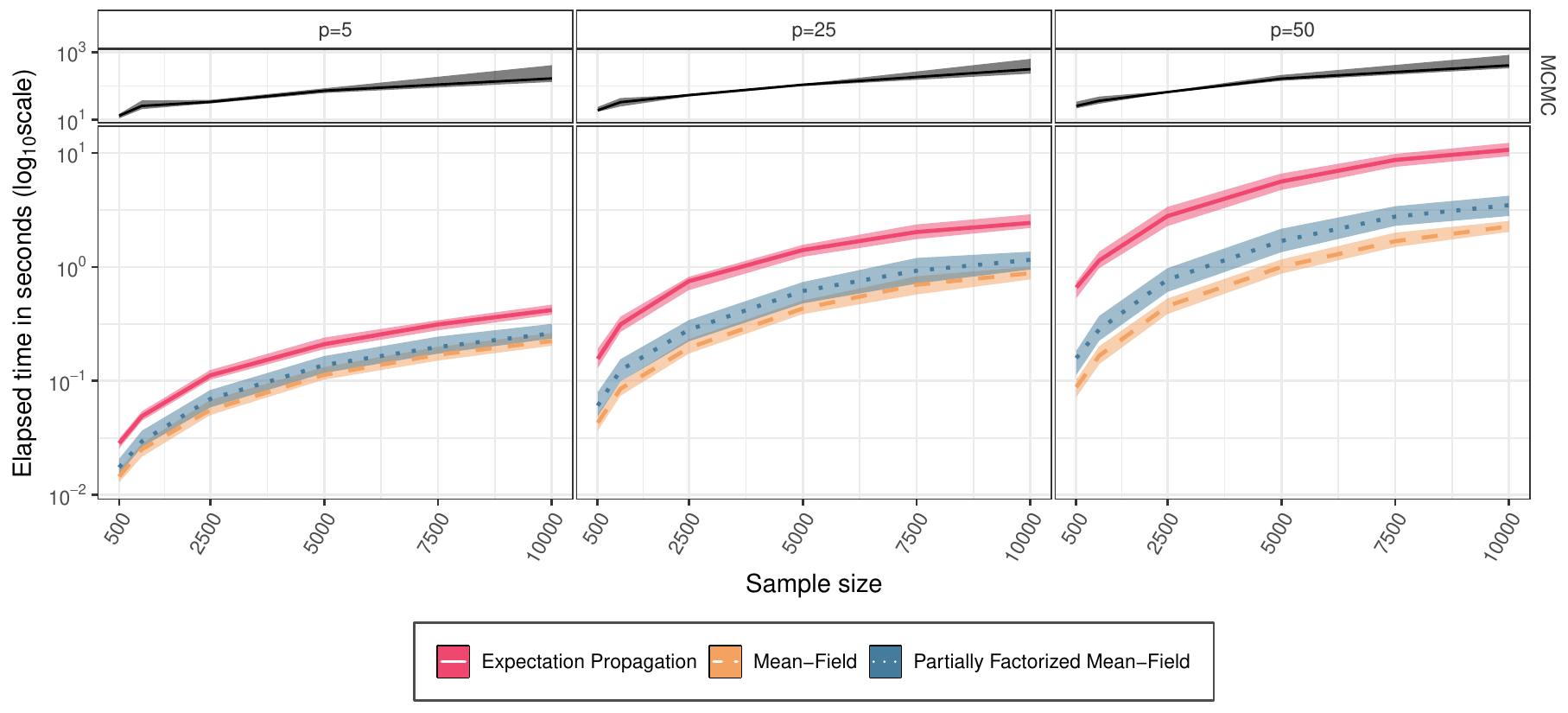}
\caption{Computational times of the proposed approximations, in seconds (log scale); lines indicate the median across $100$ replications, while the shaded areas represent the first and third quartiles. Top panels report \textsc{mcmc} elapsed time, using a narrower axis scale for improved readability.}
\label{fig:sim_time}
\end{figure}

Focusing on computational performance, Figure~\ref{fig:sim_time} compares the elapsed times of the three proposed approximations across the different simulation settings, including the estimation of the thresholds via approximate marginal likelihood optimization. 
Overall, \textsc{mfvb} emerges as the fastest routine, followed by \textsc{pmf} and \textsc{ep}. 
The implementations provided at \repo{} are specifically designed for settings with large $n$ and moderate $p$; in the intermediate scenarios with $p = 25$ covariates, the required computational time for the three approximations ranges from $0.1$ to $3$ seconds, depending on the sample size, on an Apple \textsc{m4} processor with optimized \textsc{openblas} libraries. 
For comparison, the top panels of Figure~\ref{fig:sim_time} report the elapsed time for \textsc{mcmc} across simulation settings; focusing again on the scenarios with $p=25$, collecting $5000$ \textsc{mcmc} iterations after a burnin of $1000$ requires between $20$ seconds and $5$ minutes. 
Although direct comparisons between sampling-based and optimization-based approaches should be interpreted with caution, these results clearly indicate that the proposed routines can deliver accurate posterior inference in a small fraction of the computational time required by \textsc{mcmc}.

\begin{table}[tb]
\centering
\begin{tabular}{llc@{}cccccc@{\hskip0.3\tabcolsep}cccccc@{\hskip0.3\tabcolsep}cccccc}
	&&&\multicolumn{5}{c}{$80\%$} && \multicolumn{5}{c}{$90\%$} && \multicolumn{5}{c}{$95\%$}\\
	\cmidrule{4-8}
	\cmidrule{10-14}
	\cmidrule{16-20}
	$n$	&&   & $\beta_1$ & $\beta_2$ & $\beta_3$ & $\beta_4$ & $\beta_5$ &&
		 $\beta_1$ & $\beta_2$ & $\beta_3$ & $\beta_4$ & $\beta_5$ &&
		  $\beta_1$ & $\beta_2$ & $\beta_3$ & $\beta_4$ & $\beta_5$\\
	    \toprule
$500$ & \textsc{ep}   &  & 82 & 78 & 76 & 78 & 76 && 89 & 91 & 85 & 89 & 86 && 94 & 94 & 92 & 93 & 94 \\
      & \textsc{mfvb}   &  & 72 & 66 & 67 & 68 & 70 && 83 & 83 & 80 & 80 & 78 && 90 & 91 & 85 & 91 & 85 \\
      & \textsc{pmf} &  & 80 & 73 & 73 & 74 & 74 && 88 & 91 & 85 & 86 & 84 && 93 & 91 & 90 & 93 & 93 \\
       \midrule
$1000$ & \textsc{ep}   &  & 80 & 75 & 78 & 78 & 90 && 91 & 87 & 88 & 86 & 93 && 95 & 93 & 92 & 92 & 96 \\
       & \textsc{mfvb}   &  & 73 & 62 & 73 & 72 & 80 && 82 & 82 & 84 & 80 & 92 && 90 & 88 & 89 & 85 & 94 \\
       & \textsc{pmf} &  & 80 & 71 & 76 & 77 & 88 && 87 & 87 & 87 & 86 & 93 && 93 & 91 & 91 & 89 & 96 \\
       \midrule
$2500$ & \textsc{ep}   && 86 & 80 & 80 & 80 & 79 &&90 & 86 & 91 & 90 & 86 && 96 & 95 & 98 & 94 & 92 \\
       & \textsc{mfvb} && 73 & 67 & 75 & 76 & 70 &&89 & 81 & 84 & 83 & 80 && 91 & 90 & 92 & 90 & 86 \\
       & \textsc{pmf}  && 82 & 75 & 79 & 79 & 77 &&90 & 84 & 88 & 87 & 84 && 95 & 94 & 95 & 93 & 90 \\
       \midrule
$5000$ & \textsc{ep}   &  & 80 & 84 & 84 & 76 & 79 && 90 & 93 & 89 & 88 & 90 && 92 & 95 & 92 & 94 & 94 \\
       & \textsc{mfvb}   &  & 67 & 73 & 75 & 67 & 70 && 82 & 85 & 84 & 78 & 85 && 90 & 93 & 89 & 87 & 91 \\
       & \textsc{pmf} &  & 79 & 82 & 83 & 74 & 77 && 88 & 91 & 87 & 87 & 89 && 92 & 95 & 90 & 94 & 93 \\
       \midrule
$10000$ & \textsc{ep}   &  & 79 & 78 & 86 & 82 & 80 && 91 & 91 & 87 & 92 & 90 && 95 & 96 & 96 & 98 & 94 \\
        & \textsc{mfvb}   &  & 67 & 71 & 78 & 75 & 72 && 81 & 83 & 84 & 85 & 86 && 90 & 91 & 89 & 94 & 92 \\
        & \textsc{pmf} &  & 74 & 74 & 84 & 80 & 76 && 89 & 90 & 87 & 90 & 90 && 94 & 94 & 95 & 96 & 93 \\
	  \bottomrule
\end{tabular}
\caption{Coverage of credible intervals based on the approximate posteriors; average across $100$ replications. Top row reports the nominal coverage.}
\label{tab:coverage}
\end{table}

Finally, we conclude the simulation studies by investigating the effects of potential uncertainty underestimation due to the combination of an approximation of the posterior and the use of an empirical Bayes procedure used to estimate the cutoffs.
Specifically, we focus on the settings with $p=5$ coefficients and varying sample sizes, and evaluate the frequentist coverage of credible intervals; results are reported in Table~\ref{tab:coverage}, considering Wald-type intervals for all procedures.
Consistently with Figure~\ref{fig:sim}, current findings indicate that \textsc{ep} and \textsc{pmf} intervals show empirical coverage very close to the nominal one; instead, intervals from the \textsc{mfvb} approximation tend to undercover the true parameter values even for large sample sizes.
For the Mean-Field methods, this behavior can be partly explained by comparing the analytical expressions of the covariance matrices $\bSigma^\star\mf$ and $\bSigma^\star\pmf$, reported in Algorithms~\ref{algoMFVB} and~\ref{algoPMFVB}, respectively. 
Specifically, $\bSigma^\star\mf$---the approximate posterior covariance of $\bbeta$ given $\by$ under \textsc{mfvb}---is approximated by $\bV$, the covariance of $\bbeta$ given $\by$ {\em and} $\bz$ under the true posterior~\eqref{eq:full_cond}. 
Consequently, $\bV$ is expected to be smaller (in the matrix sense) than the true marginal covariance, as conditioning on $\bz$ reduces posterior uncertainty; this behavior is consistent with the phenomenon observed in the binary probit case \citep{consonni:2007}. 
In contrast, under \textsc{pmf}, the approximate covariance $\bSigma^\star\pmf$ is obtained by marginalizing $\bz$ in~\eqref{eq:pmf_moments} with respect to its optimal approximate density $q^\star\pmf(\bz)$, which effectively adds an extra term to $\bV$ and leads to a larger, and more realistic, quantification of uncertainty.

\section{Applications}
\subsection{Brazilian bank}
\label{sec:brazil}

This section focuses on an illustrative example involving customer satisfaction for a Brazilian bank, extracted from \citet{scarpa:2012}.
Specifically, for $n=500$ customers marketing research data are available and focus on the level of satisfaction (ordinal variable with $K=4$ ordered levels) and socio-demographic information; refer to  Appendix B.3 of \citet{scarpa:2012} for a more detailed description.
We model the ordinal level of satisfaction as a function of age, a binary indicator of male gender and income (in Brazilian reais); covariates are standardized following \citet{gelman:2008}, and the regression coefficients $\bbeta=(\beta_1,\beta_2,\beta_3)\T$ are assigned a Gaussian prior with mean zero and diagonal covariance with elements equal to $2$.
The proposed approaches are compared with posterior sampling, using on the same settings as the simulations in Section~\ref{sec:sim}.

\begin{figure}[tb]
\centering
\includegraphics[width=\textwidth]{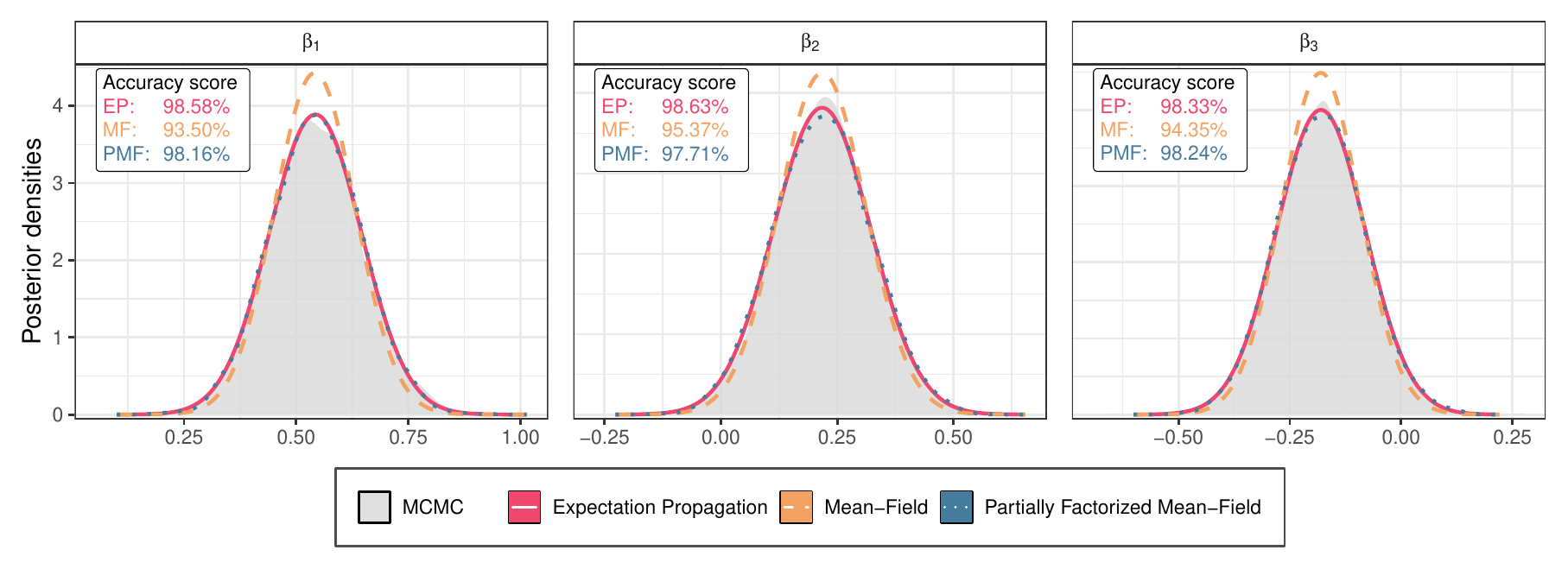}
\caption{{\em Brazilian Bank}. Univariate posterior densities estimated from \textsc{mcmc} samples (gray densities), \textsc{ep} (red curves), \textsc{mfvb} (orange curves) and \textsc{pmf} (blue curves). Accuracy scores for each method are reported in the top-left corners. }
\label{fig:brazilian}
\end{figure}

The purpose of this example is to compare the proposed approximations with posterior sampling in an illustrative case study where all methods are computational feasible in few minutes.
Figure~\ref{fig:brazilian} focuses on univariate marginals and compares sampling with the proposed posterior approximations;
$q\pmf^\star(\bbeta)$ is computed from a kernel-density estimator relying on $1000$ independent Monte Carlo samples, as described at the end of Section~\ref{sec:pmf}.
Recalling the latent regression interpretation of the cumulative probit, results indicate that older ages and being male have a positive effect and lead to higher levels of satisfaction, accounting for the other covariates; instead, income is associated with a coefficient with negative sign that suggests that increasing levels of income decrease the customer satisfaction. 
Figure~\ref{fig:brazilian} indicates that all approximations are coherent with the posterior obtained from \textsc{mcmc}; the quality in approximating the univariate marginals is measured in terms of the total variation between the ``true'' posterior $p(\beta_j\mid\by)$---as obtained from \textsc{mcmc} samples---and the corresponding approximations $q^\star(\beta_j)$, suitably rescaled as in \citet{faes:2011} to obtain the accuracy score  
\begin{equs}
	\mbox{Accuracy}_j(q^\star) &= 100\left(1 - \frac{1}{2} \int \left| p(\beta_j\mid\by) - q^\star(\beta_j)\right| \mbox{d} \beta_j\right)\%, \quad \mbox{for} \,j=1,2,3,
\end{equs}
relying on the implementation provided in the Supplementary Materials of \citet{luts:2018}. 
The accuracy scores are displayed in the top-left corners of Figure~\ref{fig:brazilian}, and indicate an exceptional performance of \textsc{ep} and \textsc{pmf}, with accuracy scores exceeding $98\%$ and $97\%$, respectively; \textsc{mfvb} scores, instead, remain satisfactory but below $95.5\%$ across all parameters, coherently with the results observed in Section~\ref{sec:sim}.

\subsection{The infinito network}
\label{sec:criminal}

In this section, we focus on a more challenging application and analyze information derived from the law enforcement operation ``{\em Operazione Infinito}'', which investigated the core structure of the 'Ndrangheta mafia in northern Italy; refer to \citet{calderoni:2017} and \citet{legramanti:2022} for more details, and to the page \url{https://sites.google.com/site/ucinetsoftware/datasets/covert-networks/ndrangheta-mafia-2} for the complete dataset.

The data used in this case study describe co-attendance at various meetings among $V = 118$ suspects identified in judicial records as potential members of the criminal organization ``\textit{La Lombardia}''. 
Pairwise relationships among generic pairs of suspects $i$ and $i^\prime$ are represented as edges of an undirected network, using an ordinal indicator $y_{i,i^\prime} \in \{1,2,3\}$.
In addition, the judicial acts provide information on each suspect's membership in local sub-units (referred to as different ``\textit{locali}'') as well as their hierarchical role within the organization \citep{legramanti:2022}.

Such relational data can be modeled via an additive social-relation regression model \citep[][Section 2.1]{hoff:2021} that includes node-specific additive effects to characterize individual variability, and dyad-specific covariates to account for membership to the same {\em locale} and holding the same role in the criminal network; this is specified letting for each edge $y_{i,i^\prime}$, $\{(i,i^\prime):1\leq i < i^\prime \leq V\}$
\begin{equation}
\begin{gathered}
\label{eq:netmod}
\mbox{pr}(y_{i,i^\prime} \leq k) = \Phi(\alpha_k - \eta_{i,i^\prime}), \quad k = 1,\ldots,K-1 \\
\eta_{i,i^\prime} = a_i + a_{i^\prime} + b^{\mbox{\scriptsize locale}}_{l} \mbox{locale}_{l, i,i^\prime} + b^{\mbox{\scriptsize role}}_{r} \mbox{role}_{r, i,i^\prime}
\end{gathered}
\end{equation}
where the parameters $a_1,\ldots,a_V$ represent node-specific effects, while the regression coefficients $(b^{\mbox{\scriptsize locale}}_1, \ldots, b^{\mbox{\scriptsize locale}}_{p_l})$ and $(b^{\mbox{\scriptsize role}}_1, \ldots, b^{\mbox{\scriptsize role}}_{p_r})$ characterize {\em locale}-specific and role-specific effects that influence the cumulative probabilities when the nodes of the dyad $(i,i^\prime)$ belong to the same {\em locale} or hold the same role in the organization, as denoted by the indicator variables
$\mbox{locale}_{l, i,i^\prime}$ and $\mbox{role}_{r, i,i^\prime}$.

The additive social-relation regression model~\eqref{eq:netmod} can be expressed as the proposed cumulative regression model~\eqref{eq:model}, letting $\bbeta = \big[(a_1,\ldots,a_V), (b^{\mbox{\scriptsize locale}}_1, \ldots, b^{\mbox{\scriptsize locale}}_{p_l}), (b^{\mbox{\scriptsize role}}_1, \ldots, b^{\mbox{\scriptsize role}}_{p_r})\big]\T$,  vectorizing the edges as $\by = (y_{2,1}, y_{3,1}, \ldots, y_{V,V-1})\T$ and constructing the design matrix $\bX$ with appropriate indicators.
Focusing on the $p_l=11$ largest {\em locali} and on the $p_r=2$ confirmed roles ({\em boss} or {\em affiliate}), the resulting specification is a model with $n=V(V-1)/2 = 6903$ observations and $p = V + p_l + p_r= 130$ covariates that can be efficiently estimated with the proposed methods.

\begin{figure}[tb]
\centering
\includegraphics[width=\textwidth]{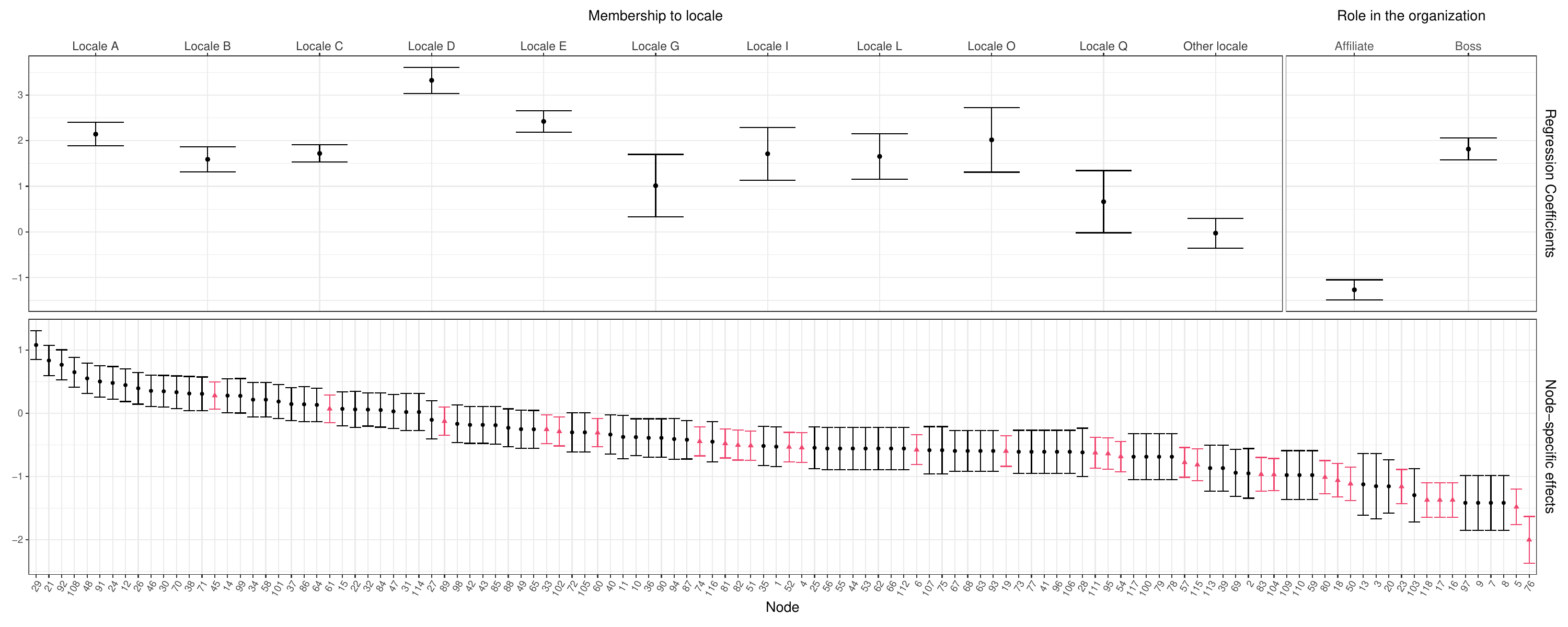}
\caption{{\em Infinito network}. Posterior distributions of the coefficients for the social-relation regression model; points denote posterior means, ticks $90\%$ credible intervals. Top panels: {\em locale}-specific and role-specific effects. Bottom panels: node-specific effects; red bars correspond to effects associated with bosses.}
\label{fig:ndrangheta}
\end{figure}

We estimate the social-relation regression model using \textsc{ep}, and the results for the various model terms are presented in Figure~\ref{fig:ndrangheta}.
The top panels display the posterior distributions of the dyad-specific effects, showing that membership in the same {\em locale} increases the cumulative probability of frequent meetings.
This finding suggests that local affiliations play a central role in shaping interactions within the network, consistent with the strong internal cohesion and coordination typically observed within mafia-type organizations \citep{calderoni:2017}; such an effect is particularly pronounced for members of {\em locali} ``D'' and ``E'', with posterior mean estimates exceeding $2$.
Interestingly, these {\em locali} represent key subgroups within the criminal organization and display distinctive interconnection patterns. 
The former (referred to as the {\em ``locale di Milano''} in the judicial documents) is recognized as one of the oldest {\em locali} and has long served as a reference point within the territory.
The latter (known as the {\em ``locale di Pioltello''}) was established more recently by affiliates who were formerly members of the {\em locale di Milano}, and its members have indeed held prominent roles in the criminal organization; further details can be found in the public documents from the Judicial Court of Milan ({\em Tribunale Ordinario di Milano, Ufficio del Giudice per le indagini preliminari}; 43733/06 R.G.N.R., 8265/06 R.G.G.I.P., available at \url{https://www.genovaweb.org/ordinanza_gip-MILANO.pdf}).
In addition, the top-right panel of Figure~\ref{fig:ndrangheta} indicates that the cumulative probability of contact increases when both individuals in the dyad are {\em bosses}, whereas it decreases when both are {\em affiliates}; this pattern seems coherent with sociological evidence suggesting that high-ranking members maintain interactions to coordinate strategic decisions and the communication flow.

The bottom panels of Figure~\ref{fig:ndrangheta} report posterior estimates for the node-specific effects, ordered by decreasing posterior means.
Results reveal substantial variability across estimated effects, which can be used to identify suspects who tend to participate more actively in meetings; for instance, the effect associated with the node labeled as ``29'' (an affiliate of {\em locale} ``L'') indicates high activity, as this individual was involved in $13$ dyads classified as ``recurrent contacts'', $26$ as ``sporadic contacts'', and only $78$ as ``absence of contacts''. 
Colored bars denote the effects of suspects identified as bosses; interestingly, most of these effects are negative: after accounting for dyad-specific relationships, the direct involvement of a boss in a generic dyad is an event characterized by lower probability. 
This result suggests that leaders maintain a pervasive but indirect control in the network, and avoid frequent exposure with minor affiliates to reduce the risk of detection.

\section{Closing remarks}
Regression models for ordinal data are popular across numerous applications; however, the scarcity of scalable algorithms for approximate Bayesian inference restricts their practical utility in large-scale settings.
This article introduces three algorithms for approximate inference in cumulative probit models, provides efficient implementations and evaluates their performance across various simulated scenarios and applications.
The results indicate that the three approaches offer complementary trade-offs between accuracy and computational cost; however, the \textsc{ep} routine emerges as the method that empirically provides the most accurate characterization of the posterior distribution.
From a theoretical perspective, \textsc{ep} procedures have fewer formal guarantees compared to methods based on Variational Bayes; for instance, the resulting algorithmic routine is not guaranteed to converge to a minimum of the reverse \textsc{kl} divergence \citep{bishop:2006}.
Nonetheless, their superior performance in numerous settings---including the ordinal probit model---motivates further research efforts to better understand the properties of such algorithms \citep[e.g.,][]{dehaene:2018}.

The proposed routines are applicable to any linear predictor specification; however, more computationally efficient variants can be developed when the design matrix shows specific sparsity structures \citep[e.g.,][Section 7]{zhou2024,gronberg:2025,goplerud:2025}.
Other potential generalizations include addressing more structured linear predictors, such as those involving latent-factor models or multiplicative effects, or extending the model to include category--specific coefficients similarly to proportional odds specifications.
Extending the proposed algorithms to these contexts is not immediate; however, the routines presented in this article might serve as a starting point for such future developments.

\section*{Acknowledgments}
The author is grateful to Daniele Durante, Augusto Fasano and Stefano Rizzelli for the fruitful suggestions during the development of this article.
The author acknowledges support from the European Union (\textsc{erc}, \textsc{nemesis}, project number: 101116718). 
Views and opinions expressed are however those of the author only and do not necessarily reflect those of the European Union or the European Research Council Executive Agency. Neither the European Union nor the granting authority can be held responsible for them.

\bibliographystyle{apalike}
\bibliography{99BIB.bib}
\newpage

\appendix %
\makeatletter
\renewcommand \thesection{S\@arabic\c@section}
\renewcommand\thetable{S\@arabic\c@table}
\renewcommand \thefigure{S\@arabic\c@figure}
\makeatother

\section*{Supplementary Materials}
\begin{center}
	\textbf{\Large Approximate Bayesian inference for cumulative probit regression models}

	Emanuele Aliverti, Department of Statistical Sciences, University of Padova
\end{center}

\section{Mean-field Variational Bayes}
\subsection{Derivation of the evidence lower bound}
\label{sec:suppMF}
According to the derivations provided in Equation~\ref{eq:elbo_def},  the \textsc{elbo} under the Mean-Field factorization can be written as
\begin{equs}
\papp(\by,q\mf) =& \mathbb{E}_{ q\mf(\bbeta,\bz)}[\log p(\by,\bz,\bbeta)]  - \mathbb{E}_{q\mf(\bbeta,\bz)}[\log q\mf(\bbeta,\bz)] \\
=& \mathbb{E}_{ q\mf(\bbeta,\bz)}[\log p(\by \mid \bz,\bbeta) + \log p(\bz\mid\bbeta) + \log p(\bbeta)] - \mathbb{E}_{q\mf(\bbeta,\bz)}[\log q\mf(\bbeta)]-  \mathbb{E}_{q\mf(\bbeta,\bz)}[\log q\mf(\bz)] \\
=& \sumin \mathbb{E}_{ q\mf(\bbeta,\bz)}[\log p(y_i\mid z_i,\bbeta)] + \sumin \mathbb{E}_{ q\mf(\bbeta,\bz)}\left[\log p(z_i\mid\bbeta)\right] +   \mathbb{E}_{ q\mf(\bbeta)}[\log p(\bbeta)] \\
 &- \mathbb{E}_{q\mf(\bbeta)}[\log q\mf(\bbeta)]-  \sumin\mathbb{E}_{q\mf(z_i)}[\log q\mf(z_i)]
\end{equs}
Denoting as $\bar{\bbeta} = \mathbb{E}_{ q\mf(\bbeta)}\left[\bbeta\right]$, the optimal density $q\mf(z_i)$ is a Truncated-Normal over the interval $[\alpha_{y_i-1}, \alpha_{y_i}]$ with location $\bxit\bar{\bbeta}$ and unit scale; therefore,
\begin{equs}
\mathbb{E}_{q\mf(z_i)}[\log q\mf(z_i)]  &= \mathbb{E}_{q\mf(z_i)} \left[\log \left\{\phi(z_i - \bxit\bar{\bbeta}, 1)\right\} \right]-\log[{\Phi(\alpha_{y_i-1} -\bxit\bar{\bbeta}) - \Phi(\alpha_{y_i} -\bxit\bar{\bbeta})}] \\
					&+ \mathbb{E}_{q\mf(z_i)} \left[\log \left\{ \mathds{I}[\alpha_{k-1} < z_i < \alpha_k]\right\} \right]\\
					&= \mathbb{E}_{q\mf(\bbeta)q\mf(z_i)} \left[\log p(z_i\mid\bbeta) \right] -\log[{\Phi(\alpha_{y_i} -\bxit\bar{\bbeta}) - \Phi(\alpha_{y_i-1} -\bxit\bar{\bbeta})}].
\end{equs}
Considering only components that depend on variational parameters
\begin{equs}
\papp(\by,q\mf) =& \sumin \log[{\Phi(\alpha_{y_i} -\bxit\bar{\bbeta}) - \Phi(\alpha_{y_i-1} -\bxit\bar{\bbeta})}] + \mathbb{E}_{ q\mf(\bbeta)}[\log p(\bbeta)] - \mathbb{E}_{q\mf(\bbeta)}[\log q\mf(\bbeta)]\\
=& \sumin \log[{\Phi(\alpha_{y_i} -\bxit\bar{\bbeta}) - \Phi(\alpha_{y_i-1} -\bxit\bar{\bbeta})}] -\frac{1}{2} (\bar{\bbeta}-\bmu_0)\T\bSigma_0^{-1}(\bar{\bbeta}-\bmu_0) + const
\end{equs}

\section{Partially factorized Mean-Field Variational Bayes}
\subsection{Derivation of Algorithm 2}
\label{supp:pmf}
\citet{holmes:2006} developed an efficient sampler for binary regression models based on data augmentation that can be generalized to the cumulative probit setting. 
As a preliminary result, note that applying the Woodbury identity it holds that
\begin{equs}
\bX\bV\bX\T &= \bX(\bX\T\bX + \bSigma_0^{-1})^{-1}\bX\T \\
	    &= \bX(\bSigma_0 - \bSigma_0\bX\T(\mathbf{I}_n + \bX\bSigma_0\bX\T)^{-1}\bX\bSigma_0)\bX\T \\
	    &= \bX\bSigma_0\bX\T - \bX\bSigma_0\bX\T(\mathbf{I}_n + \bX\bSigma_0\bX\T)^{-1}\bX\bSigma_0\bX\T \\
	    &= \bX\bSigma_0\bX\T(\mathbf{I}_n + \bX\bSigma_0\bX\T)^{-1}(\mathbf{I}_n + \bX\bSigma_0\bX\T) - \bX\bSigma_0\bX\T(\mathbf{I}_n + \bX\bSigma_0\bX\T)^{-1}\bX\bSigma_0\bX\T \\
	    &=  \bX\bSigma_0\bX\T(\mathbf{I}_n + \bX\bSigma_0\bX\T)^{-1}.
\end{equs}
As outlined in Section~\ref{sec:pmf}, according to standard Gaussian properties
\begin{equs}
	p(\bz\mid\by) = \int p(\bz\mid\by,\bbeta)p(\bbeta) \mbox{d}\bbeta \propto \phi_n(\bz -\bX\bmu_0, \mathbf{I}_n + \bX\bSigma_0\bX\T) \prod_{i=1}^n\mathds{I}[\alpha_{y_i-1} < z_i < \alpha_{y_i}].
\end{equs}
Due to the closure properties of Truncated-Normals, the conditional distribution of $(z_i\mid \bz\mi, \by)$ is a univariate Truncated-Normal with parameters that can be obtained leveraging the properties of conditional multivariate Gaussians. 
Focusing on a generic $z_i$, the joint distribution of $(z_i,\bz\mi\mid\by)$ reads as
\begin{equs}
\left(
 \begin{bmatrix}
	 z_i \\
	 \bz\mi
\end{bmatrix}\large\mid  \by\right) \sim N_n \left(
 \begin{bmatrix}
	 \bxit\bmu_0\\
	 \bX\mi\bmu_0
\end{bmatrix}, 
 \begin{bmatrix}
	 1 + \bxit\bSigma_0\bxii & \bxit\bSigma_0\bX\mi\T\\
	 \bX\mi\bSigma_0\bxii &  \mathbf{I}_{n-1} + \bX\mi\bSigma_0\bX\mi\T
\end{bmatrix}
\right).
\end{equs}
The conditional variance can be obtained relying on Sherman-Morrison as
\begin{equs}
\mbox{var}[z_i\mid\bz\mi,\by] &= 1 + \bxit\bSigma_0\bxii - \bxit\bSigma_0\bX\mi\T(\mathbf{I}_{n-1} + \bX\mi\bSigma_o\bX\mi\T)^{-1}	 \bX\mi\bSigma_0\bxii \\
			      &= 1 + \bxit\bSigma_0\bxii + \bxit\left( \left(\bX\mi\T\bX\mi + \bSigma_0^{-1}\right)^{-1} - \bSigma_0 \right)\bxii\\
			      &= 1 + \bxit\left(\bX\mi\T\bX\mi + \bSigma_0^{-1}\right)^{-1}\bxii\\
			      &= 1 + \bxit\left(\bX\T\bX -\bxii\bxit + \bSigma_0^{-1}\right)^{-1}\bxii\\
			      &= 1 + \bxit\left(\bV + \frac{\bV\bxii\bxit\bV}{1-\bxit\bV\bxii}\right)\bxii = \frac{1}{1-\bxit\bV\bxii}.
\end{equs}
Similarly, the conditional expectation is
\begin{equs}
\mathbb{E}[z_i\mid\bz\mi,\by] &= \bxit\bmu_0 + \bxit\bSigma_0\bX\mi\T(\mathbf{I}_{n-1} + \bX\mi\bSigma_0\bX\mi\T)^{-1}	\left(\bz\mi - \bX\mi\bmu_0\right).
\end{equs}
Focusing on  the term
\begin{equs}
\bxit\bSigma_0\bX\mi\T(\mathbf{I}_{n-1} + \bX\mi\bSigma_0\bX\mi\T)^{-1} & = \bxit\bSigma_0\bX\mi\T\left(\mathbf{I}_{n-1} - \bX\mi \left(\bSigma_0^{-1} + \bX\mi\T\bX\mi\right)^{-1}\bX\mi\T\right) \\
 & = \bxit\bSigma_0\bX\mi\T - \bxit\bSigma_0\bX\mi\T\bX\mi \left(\bSigma_0^{-1} + \bX\mi\T\bX\mi\right)^{-1}\bX\mi\T,
\end{equs}
the first addend can be expanded as
\begin{equs}
\bxit\bSigma_0\bX\mi\T  & = \bxit\bSigma_0\left(\bSigma_0^{-1} + \bX\mi\T\bX\mi\right)\left(\bSigma_0^{-1} + \bX\mi\T\bX\mi\right)^{-1}\bX\mi\T\\
			& = \bxit \bSigma_0 \bSigma_0^{-1}\left(\bSigma_0^{-1} + \bX\mi\T\bX\mi\right)^{-1}\bX\mi\T + \bxit\bSigma_0 \bX\mi\T\bX\mi\left(\bSigma_0^{-1} + \bX\mi\T\bX\mi\right)^{-1}\bX\mi\T\\
			& = \bxit\left(\bSigma_0^{-1} + \bX\mi\T\bX\mi\right)^{-1}\bX\mi\T + \bxit\bSigma_0 \bX\mi\T\bX\mi\left(\bSigma_0^{-1} + \bX\mi\T\bX\mi\right)^{-1}\bX\mi\T,
\end{equs}
and therefore $\bxit\bSigma_0\bX\mi\T(\mathbf{I}_{n-1} + \bX\mi\bSigma_0\bX\mi\T)^{-1} = \bxit\left(\bSigma_0^{-1} + \bX\mi\T\bX\mi\right)^{-1}\bX\mi\T$. The overall conditional expectation is expressed more compactly relying again on Woodbury identity as
\begin{equs}
\mathbb{E}[z_i\mid\bz\mi,\by] &= \bxit\bmu_0 + \bxit\bSigma_0\bX\mi\T(\mathbf{I}_{n-1} + \bX\mi\bSigma_0\bX\mi\T)^{-1}	\left(\bz\mi - \bX\mi\bmu_0\right)\\
 & = \bxit\bmu_0 + \bxit \left(\bSigma_0^{-1} + \bX\mi\T\bX\mi\right)^{-1}\bX\mi\T 	\left(\bz\mi - \bX\mi\bmu_0\right)\\ 
 & = \bxit\bmu_0 + \bxit \left(\bSigma_0^{-1} + \bX\T\bX - \bxii\bxit\right)^{-1}\bX\mi\T \left(\bz\mi - \bX\mi\bmu_0\right)\\
 & = \bxit\bmu_0 + \bxit \left(\bV + \frac{\bV\bxii\bxit\bV}{1-\bxit\bV\bxii}\right)\bX\mi\T \left(\bz\mi - \bX\mi\bmu_0\right)\\
 & = \bxit\bmu_0 + \frac{1}{1-\bxit\bV\bxii} \bxit \bV \bX\mi\T \left(\bz\mi - \bX\mi\bmu_0\right).
\end{equs}
The updates provided in Algorithm 2 are obtained noticing that for a generic $z_i$ having a Truncated-Normal density over the interval $[\alpha_{y_i-1}, \alpha_{y_i}]$ with location $\xi_i$ and scale ${\sigma_i^\star}$ it holds that
\begin{equs}
\mathbb{E}_{q\pmf(z_i)}\left[z_i\right] = \xi_i - \sigma_i^\star\zeta_1(\tilde{u}_i, \tilde{v}_i),\quad
\mbox{var}_{q\pmf(z_i)}\left[z_i\right] = {\sigma_i^\star}^2 (1 - \zeta_1(\tilde{u}_i, \tilde{v}_i)^2 - \zeta_2(\tilde{u}_i, \tilde{v}_i)),
\end{equs}
with $\tilde{u}_i = (\alpha_{y_i-1} - \xi_i)/\sigma_i^\star$, $\tilde{v}_i = (\alpha_{y_i} - \xi_i)/\sigma_i^\star$ and relying on
the univariate functions $\zeta_1(a,b) = [\phi(b) - \phi(a)][\Phi(b) - \Phi(a)]^{-1}$ and $\zeta_2(a,b) = [b\phi(b) - a\phi(a)][\Phi(b) - \Phi(a)]^{-1}$.
\subsection{Derivation of the evidence lower bound}
\label{supp:pmf_elbo}
According to Section 1 in the Supplementary Materials of \citet{fasano:2022}, the \textsc{elbo} under a partially factorized mean-field specification can be written as
\begin{equs}
\papp(\by,q\pmf) &= \mathbb{E}_{ q\pmf(\bz)}[\log p(\by,\bz)] -  \mathbb{E}_{q\pmf(\bz)}[\log q\pmf(\bz)],
\end{equs}
since according to Equation~\eqref{eq:elbo_pmf} the optimal $q\pmf^\star(\bbeta\mid\bz)=p(\bbeta\mid\bz)$. 

Leveraging Woodbury identity, ${(\mathbf{I}_n + \bX\bSigma_0\bX\T)^{-1} = (\mathbf{I}_n - \bX(\bX\T\bX + \bSigma_0^{-1})^{-1}\bX\T) = (\mathbf{I}_n - \bX\bV\bX\T)}$, and since the prior marginal density of $\bz$ is ${p(\bz) = \int p(\bz\mid\bbeta)p(\bbeta)\mbox{d}\bbeta = \phi_n(\bz-\bX\bmu_0, \mathbf{I}_n + \bX\bSigma_0\bX\T)}$ the first addend of the \textsc{elbo} is
\begin{equs}
\mathbb{E}_{q\pmf(\bz)}[\log p(\by,\bz)]&= \mathbb{E}_{q\pmf(\bz)}[\log \left\{p(\by\mid\bz)p(\bz)\right\}]\\
					&= \mathbb{E}_{q\pmf(\bz)}\left[\sumin \log \sum_{k=1}^K\mathds{I}[\alpha_{k-1} \leq z_i < \alpha_k]\cdot k\right] + \mathbb{E}_{q\pmf(\bz)}[\log p(\bz)] + const \\
					&=\mathbb{E}_{q\pmf(\bz)}\left[-\frac{1}{2}\bz\T(\mathbf{I}_n + \bX\bSigma_0\bX\T)^{-1}\bz - \bz\T(\mathbf{I}_n + \bX\bSigma_0\bX\T)^{-1}\bX\bmu_0 \right] + const\\
					&=\mathbb{E}_{q\pmf(\bz)}\left[-\frac{1}{2}\bz\T(\mathbf{I}_n - \bX\bV\bX\T)\bz + \bz\T(\mathbf{I}_n - \bX\bV\bX\T)\bX\bmu_0 \right] + const\\
					&=-\frac{1}{2} \bar{\bz}\T(\mathbf{I}_n - \bX\bV\bX\T)\bar{\bz} -\frac{1}{2}\mbox{tr}\left((\mathbf{I}_n - \bX\bV\bX\T)\bOmega\right) + \bar{\bz}\T(\mathbf{I}_n - \bX\bV\bX\T)\bX\bmu_0 + const,
\end{equs}
where $\bar{\bz} = (\bar{z}_1, \ldots, \bar{z}_n)\T$ and $\bOmega=\mbox{diag}(\omega_1,\ldots,\omega_n)$, with $\bar{z}_i=\mathbb{E}_{q\pmf(z_i)}[z_i]$ and $\omega_i = \mbox{var}_{q\pmf(z_i)}[z_i]$ for $i=1,\ldots,n$.
The second term of the \textsc{elbo} is derived recalling that $q\pmf(z_i)$ is the density of a Truncated-Normal over the interval $[\alpha_{y_i-1}, \alpha_{y_i}]$ with location $\xi_i$ and scale ${\sigma_i^\star}$; letting $\tilde{u}_i = (\alpha_{y_i-1} - \xi_i)/\sigma_i^\star$ and  $\tilde{v}_i = (\alpha_{y_i} - \xi_i)/\sigma_i^\star$,
\begin{equs}
 \mathbb{E}_{q\pmf(\bz)}[\log q\pmf(\bz)] &= \sumin \mathbb{E}_{q\pmf(z_i)}\left[\log q\pmf(z_i)\right]\\
					  &= \sumin \mathbb{E}_{q\pmf(z_i)}\left[\log \left\{\dfrac{1}{\Phi(\tilde{v}_i) - \Phi(\tilde{u}_i)}\phi(z_i - \xi_i, {\sigma_i^\star}^2)\mathds{I}[\alpha_{y_i-1} \leq z_i < \alpha_{y_i}]\right\} \right]\\
					   &= -\sumin \log({\Phi(\tilde{v}_i) - \Phi(\tilde{u}_i)}) + \sumin\mathbb{E}_{q\pmf(z_i)}\left[-\frac{1}{2{\sigma_i^\star}^2}(z_i-\xi_i)^2\right] + const\\
					   &= -\sumin \log({\Phi(\tilde{v}_i) - \Phi(\tilde{u}_i)})-\frac{1}{2}\sumin\frac{\omega_i + \bar{z}_i^2}{ {\sigma_i^\star}^2}  -\frac{1}{2}\sumin\dfrac{\xi_i^2}{{\sigma_i^\star}^2} + \sumin\frac{\bar{z}_i\xi_i}{ {\sigma_i^\star}^2}   + const.
\end{equs}
The expression of the \textsc{elbo} is obtained by combining the two addends and simplifying the term $\sumin (\omega_i + \bar{z}_i^2)/{ {\sigma_i^\star}^2}$ noticing that ${\mbox{tr}\left((\mathbf{I}_n - \bX\bV\bX\T)\bOmega\right) = \sumin \omega_i / { {\sigma_i^\star}^2}}$ since ${ {\sigma_i^\star}^2} = 1/(1-\bxit\bV\bxii)$.

\subsection{Computational considerations for large $n$}
\label{supp:pmf_comp}
In order to obtain an efficient implementation of the \textsc{pmf} routine when $n$ is large, we follow the specifications provided in the Supplementary Materials of \citet{fasano:2022}.
Indeed, the most expensive parts of Algorithm~\ref{algoPMFVB} is on computing the Truncated-Normal locations $\xi_i$; this operation is efficiently performed pre-computing the matrix $\bX\bV$ with generic row $(\bX\bV)_i =\bxit\bV$, and noticing that
\begin{equs}
\xi_i = \bxit\bmu_0 + {\sigma_i^\star}^2 \bxit \bV \bX\mi\T \left(\bar{\bz}\mi - \bX\mi\bmu_0\right)
= \bxit\bmu_0 - {\sigma_i^\star}^2 (\bX\bV)_i\bX\mi\T\bX\mi\bmu_0 + {\sigma_i^\star}^2 (\bX\bV)_i \bD_i\T.
\end{equs}
The first two scalar addends can be pre-computed, while the third depends on an $(n\times p)$ matrix $\bD$ with generic row $\bD_i=\bar{\bz}\T\bX-\bar{z}_i\bxit$, that can be updated recursively (before computing $\xi_i$) at each  iteration as $\bD_i = \bD_{i-1} + \bar{z}_{i-1}\bx_{i-1}\T - \bar{z}_{i}\bxit$. 
Similarly, the optimal scales ${\sigma_i^\star}^2$ are pre-computed from the diagonal elements of $\bX\bV\bX\T$, and such elements can be directly obtained multiplying $\bX\bV$ and $\bX$ elementwise and summing by row, without wasteful operations \citep[e.g.,][Appendix B.1]{wood:2017}; such quantities also allow to compute efficiently the \textsc{elbo}, avoiding the explicit construction and storage of $(n\times n)$ matrices.

\section{Expectation Propagation}
\subsection{The Selection-Normal distribution}
\label{appendix:selnorm}
Recalling Section 3.2 of \citet{arellano:2006}, the random vector $\mathbf{S}\in\mathbb{R}^p$ follows a Selection-Normal distribution if it is characterized by the probability density function
\begin{equs}[e:block]
\label{eq:sel_norm_pdf}
f_\bS(\mathbf{s}) &= \phi_p(\mathbf{s} - \boldsymbol{\gamma}_\bW, \bUpsilon_\bW) \frac{\bar{\Phi}_q(\mathcal{C};\bDelta\T\bUpsilon^{-1}_\bW(\mathbf{s} - \boldsymbol{\gamma}_\bW) + \boldsymbol{\gamma}_\bU, \bUpsilon_\bU - \bDelta\T\bUpsilon_\bW^{-1}\bDelta)}{\bar{\Phi}_q(\mathcal{C};\boldsymbol{\gamma}_\bU,\bUpsilon_\bU)},
\end{equs}
where $\bar{\Phi}_q(\mathcal{C}; \mathbf{a}, \mathbf{B})$ denotes the probability that a $q$-variate Gaussian with mean vector $\mathbf{a}$ and covariance $\mathbf{B}$ lies in the region $\mathcal{C}\in\mathbb{R}^q$; the kernel of the hybrid distribution provided in Equation~\eqref{eq:hybrid}
\begin{equs}
h_i(\bbeta) &= \frac{1}{Z_{h_i}}\left[\Phi(\alpha_{y_i} -\bxit\bbeta) - \Phi(\alpha_{y_i-1} -\bxit\bbeta)\right]\phi_p(\bbeta-\bQ^{-1}\mi\br\mi, \bQ^{-1}\mi)
\end{equs}
is in the same structure as~\eqref{eq:sel_norm_pdf} letting $q=1$ and
\begin{equs}[e:block]
\label{eq:sel_norm_pars}
\boldsymbol{\gamma}_\bW &= \bQ\mi^{-1}\br\mi, \quad &\bUpsilon_\bW &=\bQ\mi^{-1},\quad &\bDelta &= \bQ\mi^{-1}\bx_i, \\
\boldsymbol{\gamma}_\bU &= \bxit\bQ\mi^{-1}\br\mi, \quad &\bUpsilon_\bU &= 1+ \bxit\bQ\mi^{-1}\bx_i, \quad &\mathcal{C} &= [\alpha_{y_{i-1}}, \alpha_{y_i}].
\end{equs}
The normalizing constant $Z_{h_i}$ of the hybrid is obtained immediately: since $q=1$, the Gaussian probability over the interval $\mathcal{C} \in \mathbb{R}$ can be expressed as the difference between two univariate Gaussian \textsc{cdf}s as $Z_{h_i} = \Phi(\bar{v}_i) - \Phi(\bar{u}_i)$, letting
\begin{equs}
\bar{u}_i = \frac{\alpha_{y_i -1} - \bx_i^\intercal\Qmi^{-1}\rmi}{(1+\bx_i^\intercal\Qmi^{-1}\bx_i)^{1/2}}, \quad \bar{v}_i = \frac{\alpha_{y_i} - \bx_i^\intercal\Qmi^{-1}\rmi}{(1+\bx_i^\intercal\Qmi^{-1}\bx_i)^{1/2}}.
\end{equs}
The mean vector $\bmu_{h_i}$ and covariance matrix $\bSigma_{h_i}$ of the hybrid distribution are obtained combining the additive construction of the selection normal developed in \citet{arellano:2006} and results on univariate Truncated-Normal moments, that can be expressed through the functions $\zeta_1(\cdot,\cdot)$ and $\zeta_2(\cdot,\cdot)$ defined previously; since $q=1$, $\bUpsilon_\bU\in\mathbb{R}^+$, $\mathcal{C}=[c_1,c_2]\in\mathbb{R}$ and letting $\bar{c}_1 := \bUpsilon_\bU^{-1/2}(c_1 - \boldsymbol{\gamma}_\bU)$, $\bar{c}_2 := \bUpsilon_\bU^{-1/2}(c_2 - \boldsymbol{\gamma}_\bU)$,
\begin{equs}
\mathbb{E}_{f_\bS(\mathbf{s})}[\mathbf{S}] = \bxi_\bW - \frac{\zeta_{1}(\bar{c}_1, \bar{c}_2)}{\bUpsilon_\bU^{1/2}}\bDelta,\quad
\mbox{var}_{f_\bS(\mathbf{s})}[\mathbf{S}] = \bUpsilon_\bW - \frac{\zeta_{1}(\bar{c}_1, \bar{c}_2)^2 + \zeta_{2}(\bar{c}_1, \bar{c}_2)}{\bUpsilon_\bU}\bDelta\bDelta\T.
\end{equs}
The explicit expressions for $\bmu_{h_i}$ and $\bSigma_{h_i}$ are obtained combining~\eqref{eq:sel_norm_pars} and the previous expression, and after some simplifications 
\begin{equs}
	\bmu_{h_i} &= \Qmi^{-1} \br\mi -  \frac{\zeta_{1}(\bar{u}_i,\bar{v}_i )}{(1+\bx_i^\intercal\Qmi^{-1}\bx_i)^{1/2} } \Qmi^{-1}\bx_i,\quad
	\bSigma_{h_i} &= \Qmi^{-1} - \left(\frac{\zeta_{1}(\bar{u}_i,\bar{v}_i)^2 + \zeta_{2}(\bar{u}_i,\bar{v}_i)}{1+\bx_i^\intercal\Qmi^{-1}\bx_i}\right) \Qmi^{-1}\bx_i\bx_i^\intercal\Qmi^{-1}.
\end{equs}

\subsection{Computational details on Expectation Propagation updates}
\label{appendix:ep}
The moment matching conditions~\eqref{eq:moment_sol} require, in principle, the inverse of the matrix $\bSigma_{h_i}$ and the product $\bSigma_{h_i}^{-1}\bmu_{h_i}$; both operations are not computed directly since the inverse of $\bSigma_{h_i}$ is easily computed via Sherman-Morrison as
\begin{equs}
	\bSigma_{h_i}^{-1} &= \left(\Qmi^{-1} - \left(\frac{\zeta_{1}(\bar{u}_i,\bar{v}_i)^2 + \zeta_{2}(\bar{u}_i,\bar{v}_i)}{1+\bx_i^\intercal\Qmi^{-1}\bx_i}\right) \Qmi^{-1}\bx_i\bx_i^\intercal\Qmi^{-1}\right)^{-1}\\
			   &= \Qmi +  \left(\frac{{\zeta_{1}(\bar{u}_i,\bar{v}_i)^2 + \zeta_{2}(\bar{u}_i,\bar{v}_i)}}{ 1 + \bx_i^\intercal \Qmi^{-1} \bx_i [1 - {\zeta_{1}(\bar{u}_i,\bar{v}_i)^2 - \zeta_{2}(\bar{u}_i,\bar{v}_i)}]}\right)\bx_i\bx_i^\intercal,
\end{equs}
and combining with the moment matching conditions~\eqref{eq:moment_sol} the natural parameters for site $i$ are updated as
\begin{equs}
\bQ_i &= (\bSigma_{h_i}^{-1} - \Qmi) = \left( \frac{{\zeta_{1}(\bar{u}_i,\bar{v}_i)^2 + \zeta_{2}(\bar{u}_i,\bar{v}_i)}}{ 1 + \bx_i^\intercal \Qmi^{-1} \bx_i [1 - {\zeta_{1}(\bar{u}_i,\bar{v}_i)^2 - \zeta_{2}(\bar{u}_i,\bar{v}_i)}]}\right)\bx_i\bx_i^\intercal = k_i\,\bx_i\bx_i^\intercal, \\
\br_i &= (\bSigma_{h_i}^{-1}\bmu_{h_i} - \rmi) = \left[(\bx_i^\intercal \Qmi^{-1} \rmi)k_i - \zeta_1(\bar{u}_i, \bar{v}_i) (1+\bxit\Qmi^{-1}\bx_i)^{1/2}\right] \bx_i = w_i\,\bx_i,
\end{equs}
where the scalar quantities $(k_i,w_i)$ correspond to
\begin{equs}
k_i &= \frac{{\zeta_{1}(\bar{u}_i,\bar{v}_i)^2 + \zeta_{2}(\bar{u}_i,\bar{v}_i)}}{ 1 + \bx_i^\intercal \Qmi^{-1} \bx_i [1 - {\zeta_{1}(\bar{u}_i,\bar{v}_i)^2 - \zeta_{2}(\bar{u}_i,\bar{v}_i)}]},\\
w_i &= (\bx_i^\intercal \Qmi^{-1} \rmi)k_i - \zeta_1(\bar{u}_i, \bar{v}_i) (1+\bxit\Qmi^{-1}\bx_i)^{1/2}.
\end{equs}
The updates for the global \textsc{ep} parameters $(\Qep,\rep)$  can be directly obtained as
\begin{equs}
\Qep &= (\Qmi + \bQ_i) = \Qmi + k_i \bx_i\bx_i^\intercal,\quad \rep &= (\br\mi + \br_i) = \rmi + w_i \bx_i.
\end{equs}
Algorithm~\ref{algoEP} illustrates the \textsc{ep} routines directly in terms of covariance matrices $\bS\ep=\Qep^{-1}$ and $\bS\mi=\Qmiinv$ relying again on Sherman Morrison, since from the previous expression the cavity covariance can be expressed as
\begin{equs}
	\Smi=\Qmiinv = (\Qep - k_i \bxii\bxit)^{-1} &= \Qep^{-1} + \frac{k_i}{1-k_i \bx_i^\intercal\Qep^{-1}\bx_i} \Qep^{-1}\bx_i\bx_i^\intercal \Qep^{-1}\\
 &= \Sep + \frac{k_i}{1-k_i \bx_i^\intercal\Sep\bx_i} \Sep\bx_i\bx_i^\intercal \Sep.
\end{equs}
Similarly, all the required scalar quantities can be expressed as
\begin{equs}
	\multicol{2}{\bar{u}_i = \frac{\alpha_{y_i -1} - \bx_i^\intercal\Smi\rmi}{(1+\bx_i^\intercal\Smi\bx_i)^{1/2}}, \quad \bar{v}_i = \frac{\alpha_{y_i} - \bx_i^\intercal\Smi\rmi}{(1+\bx_i^\intercal\Smi\bx_i)^{1/2}}},\\
	k_i &= \frac{{\zeta_{1}(\bar{u}_i,\bar{v}_i)^2 + \zeta_{2}(\bar{u}_i,\bar{v}_i)}}{ 1 + \bx_i^\intercal \Smi \bx_i [1 - {\zeta_{1}(\bar{u}_i,\bar{v}_i)^2 - \zeta_{2}(\bar{u}_i,\bar{v}_i)}]},\\
	w_i &= (\bx_i^\intercal \Smi \rmi)k_i - \zeta_1(\bar{u}_i, \bar{v}_i) (1+\bxit\Smi\bx_i)^{1/2},
\end{equs}
and the global covariance becomes
\begin{equs}
	\Sep = (\Qmi + k_i\bxii\bxit)^{-1} &=  \Qmi^{-1} - \left(\frac{\zeta_{1}(\bar{u}_i,\bar{v}_i)^2 + \zeta_{2}(\bar{u}_i,\bar{v}_i)}{1+\bx_i\bx_i}\right) \Qmi^{-1}\bx_i\bx_i^\intercal\Qmi^{-1}\\
					   &= \Smi - \left(\frac{\zeta_{1}(\bar{u}_i,\bar{v}_i)^2 + \zeta_{2}(\bar{u}_i,\bar{v}_i)}{1+\bx_i^\intercal\Smi\bx_i}\right) \Smi\bx_i\bx_i^\intercal\Smi.
\end{equs}

Lastly, the update for the normalizing constant $Z_i$ required for the evaluation of the approximate marginal likelihood is performed following directly Proposition 2 of \citet{anceschi:2024EP}, since
\begin{equs}
	\log Z_i  =&\log \Psi(\bQ_i + \Qmi, \br_i + \br\mi) - \log \Psi(\Qmi, \br\mi) - \log Z_{h_i} \\
		  =&\frac{1}{2}(\br_i + \br\mi)\T(\bQ_i + \Qmi)^{-1}(\br_i + \br\mi) + \frac{p}{2}\log(2\pi) - \frac{1}{2}\log|\bQ_i + \Qmi|\\
		  &- \frac{1}{2}\br\mi\T\bQ\mi^{-1}\br\mi - \frac{p}{2}\log(2\pi) + \frac{1}{2}\log|\bQ\mi| - \log Z_{h_i} \\
		  =&\frac{1}{2}(w_i\bx_i + \br\mi)\T(k_i\bx_i\bxit + \Qmi)^{-1}(w_i \bx_i + \br\mi) - \frac{1}{2}\log|w_i\bx_i\bxit + \Qmi|\\
		  &- \frac{1}{2}\br\mi\T\bQ\mi^{-1}\br\mi  + \frac{1}{2}\log|\bQ\mi| - \log Z_{h_i}\\
		  =& \frac{1}{1+k_i \bxit\Qmi^{-1}\bxii}\left[2 w_i \bxit\Qmi^{-1}\rmi + w_i^2 \bxit \Qmi^{-1}\bxii - k_i (\rmi\T\Qmi^{-1}\bxii)^2\right]\\
		  =&  \frac{1}{1+k_i \bxit\Smi\bxii}\left[2 w_i \bxit\Smi\rmi + w_i^2 \bxit \Smi\bxii - k_i (\rmi\T\Smi\bxii)^2\right].
\end{equs}

\end{document}